\pdfoutput=1
\documentclass[utf8]{frontiersSCNS} % for Science, Engineering and Humanities and Social Sciences articles
\usepackage{url,hyperref,lineno,microtype,subcaption}
\usepackage[onehalfspacing]{setspace}

\usepackage{units}
\usepackage{graphicx}
\usepackage{tabularx}
\usepackage{booktabs}

\newcommand{\say}[2]{``#1'' (P#2)}
\newcommand{\says}[1]{``#1''}
\def\keyFont{\fontsize{8}{11}\helveticabold }
\def\firstAuthorLast{Bonfert {et~al.}} %use et al only if is more than 1 author
\def\Authors{Michael Bonfert\,$^{1,*}$, Anke V. Reinschluessel\,$^{1}$, Susanne Putze\,$^{1}$, Yenchin Lai\,$^{1}$, Dmitry Alexandrovsky\,$^{1}$, Rainer Malaka\,$^{1}$ and Tanja Döring\,$^{1}$}

\def\Address{$^{1}$Digital Media Lab, University of Bremen, Bremen, Germany} %\\
\def\corrAuthor{University of Bremen, Bibliothekstr. 5, 28359 Bremen, Germany}

\def\corrEmail{bonfert@uni-bremen.de}

\begin{document}
\onecolumn
\firstpage{1}

\title[``Seeing the Faces Is So Important'' -- Team Meetings in VR]{``Seeing the Faces Is So Important'' -- Experiences From Online Team Meetings on Commercial Virtual Reality Platforms} 

\author[\firstAuthorLast ]{\Authors} %This field will be automatically populated
\address{} %This field will be automatically populated
\correspondance{} %This field will be automatically populated

\extraAuth{}% If there are more than 1 corresponding author, comment this line and uncomment the next one.
%\extraAuth{corresponding Author2 \\ Laboratory X2, Institute X2, Department X2, Organization X2, Street X2, City X2 , State XX2 (only USA, Canada and Australia), Zip Code2, X2 Country X2, email2@uni2.edu}

\maketitle

\begin{abstract}
During the Covid-19 pandemic, online meetings became common for daily teamwork in the home office. To understand the opportunities and challenges of meeting in virtual reality (VR) compared to videoconferences, we conducted the weekly team meetings of our human-computer interaction research lab on five off-the-shelf online meeting platforms over four months. After each of the 12 meetings, we asked the participants ($N~=~32$) to share their experiences, resulting in 200 completed online questionnaires.
We evaluated the ratings of the overall meeting experience and conducted an exploratory factor analysis of the quantitative data to compare VR meetings and video calls in terms of meeting involvement and co-presence. In addition, a thematic analysis of the qualitative data revealed genuine insights covering five themes: spatial aspects, meeting atmosphere, expression of emotions, meeting productivity, and user needs. We reflect on our findings gained under authentic working conditions, derive lessons learned for running successful team meetings in VR supporting different kinds of meeting formats, and discuss the team's long-term platform choice.
\tiny
 \keyFont{ \section{Keywords:} CSCW, Virtual reality, Social VR, Remote collaboration, Virtual meetings, Video conferencing, Autoethnography, Case study} 
 %All article types: you may provide up to 8 keywords; at least 5 are mandatory.
\end{abstract}

\begin{figure}[!ht]
\begin{center}
\includegraphics[width=\textwidth]{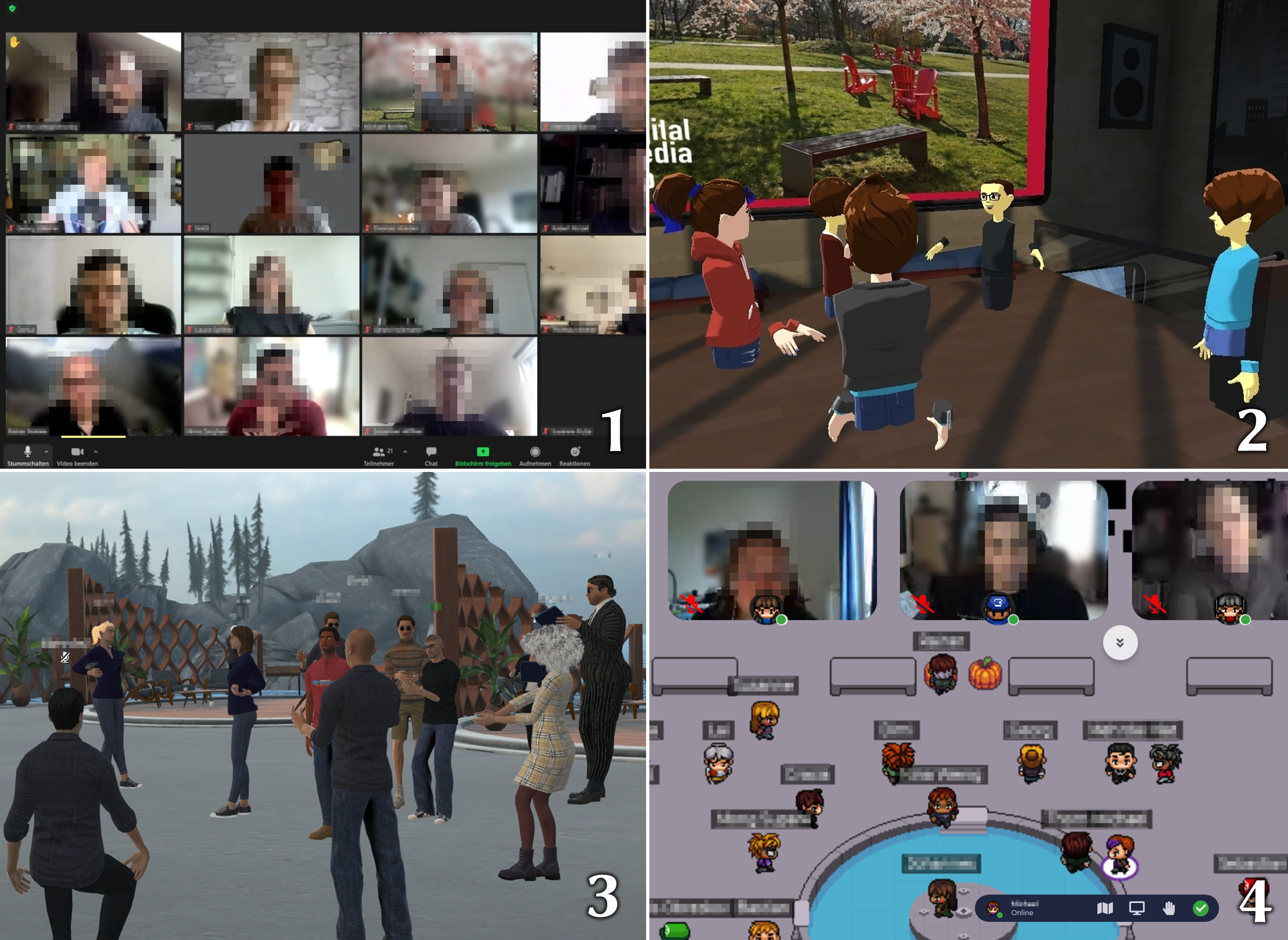}
\end{center}
\caption{The off-the-shelf platforms compared in the wild for our weekly team meetings: (1)~Our status quo, Zoom, compared to the two virtual reality platforms (2)~AltspaceVR and (3)~Engage, as well as the hybrid (4)~Gather Town combining video feeds with a spatial environment. StarLeaf was also used in the case study but is not shown here as it has a similar interface to Zoom.}
\label{fig:teaser}
\end{figure}

%%%%%%%%%%%%%%%%%%%%%%%%%%%%%%%%%%%%%%%%%%%%%%%%%%%%%%%%%%%%%%%%%
%%%%%%%%%%%%% BEGIN OF BODY %%%%%%%%%%%%%%%%%%%%%%%%%%%%%%%%%%%%%
%%%%%%%%%%%%%%%%%%%%%%%%%%%%%%%%%%%%%%%%%%%%%%%%%%%%%%%%%%%%%%%%%

\section{Introduction}
The Covid-19 pandemic required many people to work from the home office. Online meetings with video conferencing software became omnipresent after this trend had been promised for decades \citep{nilles_telecommunications_1975}.
Although video calls provide many advantages, such as seeing meeting participants or allowing for screen sharing, they still yield limitations, such as restricted social interaction between participants. 
Social virtual reality (VR) platforms might provide beneficial alternatives for online team meetings by gathering everyone in one virtual room. Previous studies showed positive psychological effects of social VR platforms \citep{Barreda-Angeles2022.PsychoEffectsSocialVRPresence} and that group behaviors and emotional responses to it are largely similar to face-to-face encounters \citep{moustafa2018.longitudinalSocialVRGroups}. However, it seems that the generation of ideas as a creative process is hindered \citep{Brucks2022ve}.  %hampers idea generation because it focuses communicators on a screen, which prompts a narrower cognitive focus. 
Beyond the faithful reproduction of in-person meetings, research further explores the possibilities of VR for enhancing social encounters and collaboration outside the restrictions of reality \citep{slater_enhancing_2016, mcveigh-schultz_case_2021}. 
\footnote{This article has been published at \href{https://www.frontiersin.org/research-topics/20749/everyday-virtual-and-augmented-reality-methods-and-applications-volume-ii\#overview}{Frontiers in Virtual Reality, \textit{Everyday Virtual and Augmented Reality: Methods and Applications, Volume II}}. Please find the final article and citation at \url{https://doi.org/10.3389/frvir.2022.945791}}

% Metaverse and hybrid work
With an increasing number of companies investing in immersive technologies and the future of the metaverse, social encounters in virtual environments (VE) will become commonplace in the foreseeable future. We can already observe this trend on platforms for personal use where users meet in VR to socialize, play, and experience cultural events together \citep{sykownik2021.ActivitiesMotivesOfUsers} -- even more pronounced during the pandemic \citep{rzeszewski2020.VRChatQuarantine}. 
% Add stats from MS Work Trend Index 2022: https://www.microsoft.com/en-us/worklab/work-trend-index/great-expectations-making-hybrid-work-work
This development also extends to the business context. In 2022, $52\%$ of employees were open to having meetings or team activities in the metaverse \citep{microsoft2022.WorkTrendIndex}. The real estate company eXp Realty is operated entirely in virtual offices with more than $75,000$ employees\footnote{\url{https://www.virbela.com/customer-stories/exp-realty}}. Numerous small companies and start-ups have launched collaborative platforms for meetings in virtual space. Also, large companies are preparing for a future in which professionals get together in virtual or blended realities, such as with Microsoft Mesh\footnote{\url{https://news.microsoft.com/innovation-stories/microsoft-mesh}} or Horizon Workrooms by Meta\footnote{\url{https://about.fb.com/news/2021/08/introducing-horizon-workrooms-remote-collaboration-reimagined}}.
Enabled by rapid technological development, immersive teamwork is more topical than ever. Still, many advertised features are visions of the future and blur the public perception of present possibilities. %How far are we on the path to remote collaboration in VEs?

% Our study
To explore current opportunities and restrictions of commercial VR meeting software and to gain a deeper first-hand understanding of the advantages and drawbacks of authentic meetings in VR, we conducted the regular team meetings of our university's human-computer interaction (HCI) research lab in VR to compare the attendees' experiences with meetings on video conferencing platforms. This experimental shift from video-only platforms was intrinsically motivated, not by conducting this study. Therefore, the authors had the opportunity to independently evaluate the experiment, resulting in genuine insights from experiences in the wild. 
Over four months in 2020, we conducted twelve meetings on five different platforms -- seven in VR and five on video conferencing or hybrid platforms for comparison. The different platforms are shown in \autoref{fig:teaser}. We accompanied the team members from the expectations before the first VR meeting to the final conclusions on the long-term software choice by collecting feedback right after each meeting through online surveys. As part of the lab team, the authors also attended the meetings, which enabled an additional autoethnographic perspective to reflect on the meetings with different platforms.

% Relating to the body of literature
With this, our case study adds to the growing body of research that investigates the gathering of people on various professional occasions on VR platforms. Previous studies explored attendees' experiences at professional social events in VR, e.g., at academic conferences and workshops \citep{krieger2021.SocialEventinVR, erickson2011.VRconferenceInteractions, lahlou2021.BeyondBeingThere, Williamson2021.ProxemicsSocialVRWorkshop}, and group dynamics in social VR outside professional context \citep{moustafa2018.longitudinalSocialVRGroups, scavarelli2021.SocialLearningSpaces}.
However, in contrast to the existing body of work, this case study focuses on the participants' personal experiences during regular online team meetings in different mediums over an extended period. Thereby, it covers both VR and video conferencing solutions. Our study aims to explore whether currently available off-the-shelf social VR platforms meet the needs and preferences of the team members for attending weekly lab meetings as a competitive alternative to conventional video conferencing solutions. Based on 200 completed survey responses, we analyze ratings and statements of the meeting experiences and the different aspects that influenced these experiences. Our analysis identifies central themes that matter for VR meetings and provides unique insights into the experiences of meeting on various platforms. While most of the individual themes have previously been discussed in the literature as isolated elements, encountering them integrated into authentic experiences of participants who compare it to videoconferencing in an in-the-wild study allowed us to assess and reconsider various factors, identify priorities, and add new aspects.

\section{Related Work}

Research on computer-mediated collaboration has a long history. A broad landscape of literature provides an understanding of how people can work together remotely comprising various modalities, cultures, technologies, and goals of remote work \citep{galegher_intellectual-teamwork_2013, raghuram_virtualWorkBridging_2019}. Video-based meeting solutions received much attention in the late 1990s \citep{finn_video-mediated_1997, hinds_cognitive_1999} and provided strong indications that video-mediated meetings can have an equally good quality as face-to-face meetings \citep{olson_face-to-face_1997}.
Still, for decades, videoconferencing was no competitive substitute for in-person meetings for large parts of the general public -- or not explored as one -- until an enforced shift due to the Covid-19 pandemic \citep{oecd_teleworking_2021}.

Along with the rise of videoconferencing platforms, the term ``Zoom fatigue'' was prominent in public and scientific discourse \citep{shockley2021.ZoomFatigue}. It describes the exhaustion after holding many or long meetings as videoconferences. Possible causes are suspected to be the cognitive load, always seeing oneself, and reduced mobility during video calls \citep{bailenson_nonverbal_2021}, but also spatial reduction of the conversation partners and their background to a 2D projection \citep{nadler_understanding_2020} calling for mitigation or alternatives. According to \cite{shockley2021.ZoomFatigue}, group belongingness was found to be the most consistent factor protecting from videoconference fatigue. Correspondingly,  early research suggested that in remote collaboration, the employed technologies impact the communication outcome dependent on the interpersonal interactivity \citep{burgoon_testing_1999}. 
While collaborating via videoconferences has been compared to other modalities, such as audio-only or in-person meetings \citep{ochsman_effects_1974, burgoon_testing_1999, daft_organizational_1986, hinds_cognitive_1999, bailenson_nonverbal_2021}, this work adds to the growing body of research with direct comparison to collaboration in VR.
The following two sections will describe previous work on characteristics of social VR that might foster interpersonal interactivity and the feeling of group belongingness or mitigate limitations of videoconferencing in other ways. We further discuss how VR technology has previously been used for social encounters in a professional context.

\subsection{Social Interactions in VR}
\label{sec:RWinteractionsInSVR}
People use social VR platforms for a broad range of reasons. The strongest motives are meeting people, staying in contact, and experiencing social presence \citep{sykownik2021.ActivitiesMotivesOfUsers}. Interacting in VEs offers considerable potential for high social presence \citep{short1976SocialPresenceOrg} as the spatial nature of the medium allows users to encounter other people in a shared space and affords complex social interactions and group dynamics. Social presence as ``the connection of people via telecommunication systems'' has been a central concept for comparing various forms of computer-mediated collaboration for decades \citep{nowak_Social-Co-Presence_2001}. Multiple factors affect how people interact with each other in VEs, including their visual and behavioral (self-)representation, the perceived agency of others' avatars, and potentially immersion \citep{kyrlitsias_social_2022}. 
Different studies yield inconclusive results on the effect of visual realism on social presence \citep{zibrek2019.PhotorealismSocialPresence,kang_impactAvatarRealism_2013, bente_noImpactAvatarRealism_2008, Nowak2003.SocialPresence} and imply dependence on other aspects of the simulation such as behavioral realism \citep{bailenson_independent_2005}, which is generally a powerful predictor of social presence \citep{oh_systematicSocialPresence_2018}, such as turn-taking in a conversation \citep{bailenson_transformed_2004}. The social VR platforms tested in this case study differ considerably in the realism of virtual human representations introducing an interesting testbed with variance.

Further, sound is an important consideration for social interactions in VR. Spatial audio has been shown to impact the user's sense of presence \citep{poeschl_SpatialSound_2013, kern_VRaudio_2020}. Although it allows the user to identify the direction of different audio sources, concurrent talking of several people poses challenges for listeners in immersive environments. It can be effectively mitigated by helpful visual cues \citep{gonzalez-franco_concurrentTalking_2017}. Besides the representation of virtual humans and verbal conversations, social VR interactions are influenced using non-verbal communication. On various social VR platforms, previous studies investigated the role of non-verbal cues and limitations in essential aspects of it and discussed similarities or discrepancies to non-verbal behavior patterns offline \citep{tanenbaum2020.ExpressiveNonverbalCommunicationinSVR, maloney_talking_2020, wigham_non-Verbal-Second-Life_2013, yee_Nonverbal-Norms_2007}. For instance, group arrangements, proxemics, and the preservation of personal space have been found to resemble offline behavior to a large extent \citep{Williamson2021.ProxemicsSocialVRWorkshop, hecht_shapePersonalSpace_2019, bailenson_equilibriumPersonalSpace_2001, yee_Nonverbal-Norms_2007} with adequate emotional responses to it \citep{wilcox_personalSpaceEmotions_2006}.
Also, previous research has shown that people can demonstrate similar social responses in virtual simulations as in reality. For example, in a study investigating compliance as done in the experiment by \cite{milgram_obedience_1978}, the participants were equally or more compliant in VR compared to a control condition in physical reality \citep{dzardanova_virtual_2021}. Unfortunately, a problem that transfers from offline social interactions to virtuality, as well, is harassment \citep{Freeman2022.VRHarassment}. As the communities are still establishing social norms for VR interactions, their governance remains a challenge \citep{blackwell_harassment_2019, blackwell_harassment-governance_2019, mcveigh-schultz_pro-social_2019}. 

Altogether, many factors determine how a person in social VR perceives and interacts with another virtual human. For the collaboration of whole teams, however, not only the interaction between two people must be considered. Related work explored group dynamics and team communication in social environments that can resemble but also go beyond the possibilities of video conferencing. 
\cite{torro_six_2021} describe the impact of non-verbal communication and spatial information of social VR and why this makes it a game-changing medium for organizations. Due to the possibility of simulating any imaginable communication process, the authors argue that social VR has the potential to exceed the communication effectiveness of video conferencing and real-world settings. As one reason, they state how new forms of group dynamics can be facilitated and how teams benefit from formal and informal encounters in VR~\citep{torro_six_2021}.

\subsection{Meeting in VR}
Researchers have explored the opportunities of remote meetings and social gatherings for a long time. Although the technical capabilities looked entirely different in 1999, the outcome of a study by \citeauthor{Stahl1999.ancientVRMeetings} still shows similarities to current work. The most significant needs concerned the feedback on the users' connection in terms of visual attention, audibility, and network connectivity. In recent years, due to technological advancements and being incentivized by Covid-19, the body of literature on meetings and events in social VR is growing rapidly. 
Typically, the study designs in this research field are not lab-based and less controlled but of observatory nature and in the wild. 

% GROUPS and DYNAMICS
In this manner, \cite{moustafa2018.longitudinalSocialVRGroups} conducted a longitudinal study on social interactions in small groups already known to each other moving their contact to social VR. The authors found that the participants experienced similar emotional states as in real-life socializing, which suggests high co-presence and transferability of existing group dynamics.
% Education
Concerning educational purposes, a literature review explored interactions relevant in social xReality (XR) learning spaces and provided an overview \citep{scavarelli2021.SocialLearningSpaces}. For example, a study by \cite{yoshimura_study_2021} reports on class meetings experienced in VR comparing access with a head-mounted display (HMD) and in desktop mode. The students attended lectures and presented in VR. The experiences highly depended on how comfortable the attendees were with the HMD and if it made them sick. 
In another experiment, \cite{ginkel2019.VRPresentationTraining} found a close resemblance between learning outcomes from training presentations in VR compared to face-to-face training. 

% Workshops and social events at conferences
Previous studies also explored the use of social VR environments in the academic community, such as social events at scientific conferences. While paper presentations can be effectively realized in videoconferences, it is much more challenging to enable virtual conference attendees to meet and connect with other participants. Yet, the informal exchange during coffee breaks and at receptions is essential to the success of academic collaborations. Therefore, organizers searched for virtual compensation, which was sometimes accompanied by scientific evaluation. 
For instance, \cite{krieger2021.SocialEventinVR} organized a virtual conference where they held a reception on the VR platform Engage and performed a qualitative evaluation. The participants reported lively experiences but struggled with the spatial audio and discomfort from the HMDs. Further, the participants felt restricted in getting to know people they had never met before due to missing facial expressions. For fostering dynamic group conversations on similar occasions, \cite{rogers2018.virtualCocktailPartyWordclouds} proposed displaying word clouds around groups hinting at the discussed topics to help strolling participants find a suitable group to join. 
Research by \cite{Williamson2021.ProxemicsSocialVRWorkshop} analyzed the user proxemics during an academic workshop in VR. The results showed proxemic interactions between attendees that are congruent to physical settings and afford dynamic group formations dependent on properties of the VE. Beyond characteristics of personal space in social interactions known from reality, the possibility to enable participants to fly with their avatar added a dimension of user proxemics that was most notable during presentation situations.

A few studies investigated the social interactions at entire conferences in VR. The first evaluation of a virtual avatar-based conference by \cite{erickson2011.VRconferenceInteractions} comprised 500 attendees in Second Life\footnote{\url{https://secondlife.com}} and was considered a ``reasonably successful'' event. According to the authors, the system must afford the formation of small groups for having focused interactions while breaking up and remixing into other groups over time. The analysis showed that the loud spatial audio disrupted conversation privacy and led to increased distances between groups which inhibited remixing. Structured social events worked better for the participants than more informal settings. In the end, none of the interviewed attendees experienced the virtual substitute and the face-to-face conference.
A decade later, \cite{lahlou2021.BeyondBeingThere} studied a conference in VR accompanied by video calls. The authors state two goals that a VR installation must meet for successful conferencing: (1) the development of knowledge and (2) informal social interaction. They find that current solutions still lack opportunities for natural social encounters and relational space. The researchers recommend careful preparation for organizers and suggest special consideration of onboarding processes and catering for socializing. 
In the same year, the conference IEEE VR 2020 was held entirely virtually for the first time. The accompanying evaluation by \cite{ahn_ieeevr2020_2021} provides a detailed comparison between the usage of different media platforms and their appropriateness for typical conference tasks. Again, the results point out the social constraints of the VE but also show advantages for connecting compared to other text-based platforms. Here, the social VR platform was rated as most appropriate for socializing and networking and providing the highest social presence. Most attendees who joined the VE decided to access it via desktop computer, although many owned an HMD. While the authors advocate for making use of the unique possibilities of virtual conferencing, they echo related research by advising caution not to transfer the substituted offline event directly to virtual space but to adapt the format and purpose of the event. 
In this spirit, \cite{mcveigh-schultz_beyond_2021} argue for conceiving solutions of VR collaboration that deliberately deviate from direct replication of familiar social encounters. Instead of imitating a face-to-face work environment, the authors promote using the full potential of immersive technologies to create an enriched experience of remote collaboration. 
The approach taken in this case study provides a flexible setting for this as the attendees could test how different platforms can serve the purposes of the meetings and reinvent the format along the way. Without predetermined system choices and autoethnographic insights, the team could adjust the conduct of the meeting every week over several months and compare the experiences. 

During the Covid-19 pandemic, researchers were challenged with finding appropriate and practical scientific methods to continue conducting studies \citep{Nind2021.ResearchPracticesCovid}.
Research on computer-mediated communication has established various methods and tools for evaluating specific aspects of the behaviors and opinions of interlocutors, e.g., turn-taking and behavioral analyses to examine meeting dynamics \citep{Samrose2021.MeetingCoach, Samrose2018.CollaborationCoach} and simulated conversations \citep{abdullah2021.videoconferenceVR}, audio-visual capturing to understand communication characteristics \citep{Koseki2020.CommProblems, byun2011.HonestSignals}, linguistic analyses \citep{fagersten2010.DiscourseAnalysis, Kramer2006.LinguisticPresence}, self-reports in interviews \citep{bleakley2022.SocialTalk} and standardized questionnaires \citep{Nowak2003.SocialPresence}, autoethnographic methods \citep{Mack2021.AutoethnographicInterns}, or a self-hosted and modified version of a social VR platform to understand the proxemics \citep{Williamson2021.ProxemicsSocialVRWorkshop}.
For this research, we rely on the conjunction of quantitative and qualitative methods from self-reports in a survey that includes standardized questionnaire items and open questions. Our approach allows longitudinal analysis through unique identifiers while preserving anonymity. 
Previous literature investigated the use of VR technology for conducting conferences or socializing events, focusing on the comparison to the face-to-face equivalent. On the other hand, this work explores the potential of holding team meetings in VR and compares it to video conferencing while investigating, in both cases, the similarity to face-to-face meetings. Instead of determining usability issues of individual social VR platforms, as previous research has done systematically \citep{liu2021.UsabilitySocialVR}, the personal experiences and needs of the meeting participants are of central interest in this case study.

% Why are we different?
  % format: meeting in particular
  % Regelmäßigkeit der Treffen
  % autoethnographic approach, flexibility, and adaptation
  % comparison to other mediums: VR vs. videoconferencing 

\section{Method}

Over four months, from August until December 2020, we evaluated twelve weekly meetings on different platforms (see Table~\ref{table:ListOfMeetings} for an overview) by inviting the participants to fill out a questionnaire after each session. We used a mixed-method approach and collected quantitative and qualitative data to investigate the participants' ratings of each meeting experience and their personal impressions. 

\begin{table}
\centering
\caption{Overview over the $12$ evaluated meetings.}
\label{table:ListOfMeetings}
\begin{tabular}{llrrrrl}
\toprule
 ID &      Medium & Participants & Responses & HMD Users & Duration (min) & Presentations \\
\midrule
 M1 &    Starleaf &                     20 &                18 & n/a       & 41 &             No \\
 M2 &    Altspace &                     19 &                16 & $87.5 \%$ & 59 &            Yes, one \\
 M3 &    Altspace &                     22 &                19 & $84.2 \%$ & 53 &            Yes, one \\
 M4 &    Altspace &                     20 &                19 & $84.2 \%$ & 21 &             No \\
 M5 &    Altspace &                     22 &                18 & $72.2 \%$ & 21 &             No \\
 M6 &    Altspace &                     20 &                16 & $68.8 \%$ & 42 &            Yes, one \\
 M7 &      Engage &                     18 &                15 & $93.3 \%$ & 57 &        Yes, three \\
 M8 &      Engage &                     15 &                11 & $72.7 \%$ & 18 &             No \\
 M9 &        Zoom &                     20 &                17 & n/a       & 43 &             No \\
M10 &        Zoom &                     24 &                18 & n/a       & 59 &            Yes, one \\
M11 & Gather Town &                     22 &                18 & n/a       & 84 &            Yes, one \\
M12 &        Zoom &                     17 &                15 & n/a       & 83 &             No \\
\bottomrule
\end{tabular}
\end{table}

\subsection{Participants} 

Generally, the weekly meeting is attended by all team members, including technical and administrative staff, undergraduate and Ph.D. students, postdocs, professors, and guests. This leads to a heterogeneous sample concerning technical literacy, previous experience with VR, and expectations towards the meetings. Participation was voluntary and strictly anonymous, so we did not collect any demographic data. Nevertheless, the participants rated their prior experience using three items: How often they use VR in general, how often they use multi-user VR applications, and whether they use VR as part of their work. From these three items, we calculated an overall prior experience score (min 0 to 10 max). On average, the $18$ participants of meeting M1 had a prior experience score of $3.167$ ($SD{=}2.41$). All group members had access to VR hardware, i.e., HTC Vive (Pro), Valve Index, or Oculus Quest 1. In total, we had $239$ meeting participations in the $12$ meetings with $32$ distinct attendees completing at least one questionnaire. $4$ participants attended and filled out the questionnaires for all $12$ meetings. $12$ participants responded after at least $10$ meetings. In each meeting were between 15 and 24 attendees. 
We did not collect demographic data such as age or gender because this would disclose the participants' identities due to the small group size. For the same reason, we did not ask about the HMD model used.

\subsection{Evaluated Meetings} 
The main purpose of the weekly team meetings was to report on and discuss current matters concerning the lab or its associates, which the group manager and administrative staff mostly did. Additionally, there were presentations by undergraduates and group members (in $\unit[50]{\%}$ of the reported meetings, at least once on each platform). At the end of the meetings, individuals and small groups would discuss matters with the group manager irrelevant to the whole team, while everybody else would already leave. On average, the meetings investigated in this study lasted $48.42$ ($SD{=}22.02$) minutes and had $19.91$ ($SD{=}2.47$) attendees with details shown in Table~\ref{table:ListOfMeetings}. 
Before the Covid-19 pandemic, the lab meetings were held in person. Between March 2020 and the beginning of this case study, they were on Zoom or StarLeaf.

\subsection{Procedure and Questionnaire} 
The meeting was joined by the attendees' device of choice -- an HMD (if applicable) or desktop client -- and followed its regular structure. During the weekly meetings, the group discussed which platform to try next and when to change it. After each meeting, the researchers emailed all group members and guests linking to a Google Forms questionnaire. Each recipient, including the authors, could individually decide to take part in the survey or not, which resulted in an average response rate of $\unit[83.7]{\%}$ (SD=$\unit[6]{\%}$). We made every effort to ensure there was no social pressure to participate and no fear of negative consequences for answering honestly.  
The questionnaire started with a consent form and a unique identifier (ID) chosen by each participant to identify repeated participation and allow longitudinal analysis anonymously. We asked how the meeting was accessed and found that HMDs were used in $92$ out of $114$ VR platform cases ($\unit[80.7]{\%}$). For the first session, we also assessed the prior experience with VR and expectations for the VR sessions.% 
The questionnaire incorporated questions from the Social Presence Questionnaire by \cite{Nowak2003.SocialPresence}, the User Burden Scale \citep{suh2016.UserBurdenScale} and self-designed questions to answer on a 7-point Likert scale. Additionally, we asked open questions about the meeting experience, e.g., group interactions, comparisons to the other platforms, and memorable experiences. The questionnaire after the first and last session had a few additional items. After the first meeting, we asked about the participants' general usage of VR, their usage of multi-user VR applications, whether they work with VR, and their expectations and worries for VR meetings. Additionally, the final survey after the last session incorporated questions about the preferred meeting platform and the reasons for the preference. In the supplemental materials, we provide an overview of all questionnaire items and, if applicable, their respective origins from standardized instruments.

\subsection{Platforms}  
\begin{figure}[!ht]
\begin{center}
\includegraphics[width=13cm]{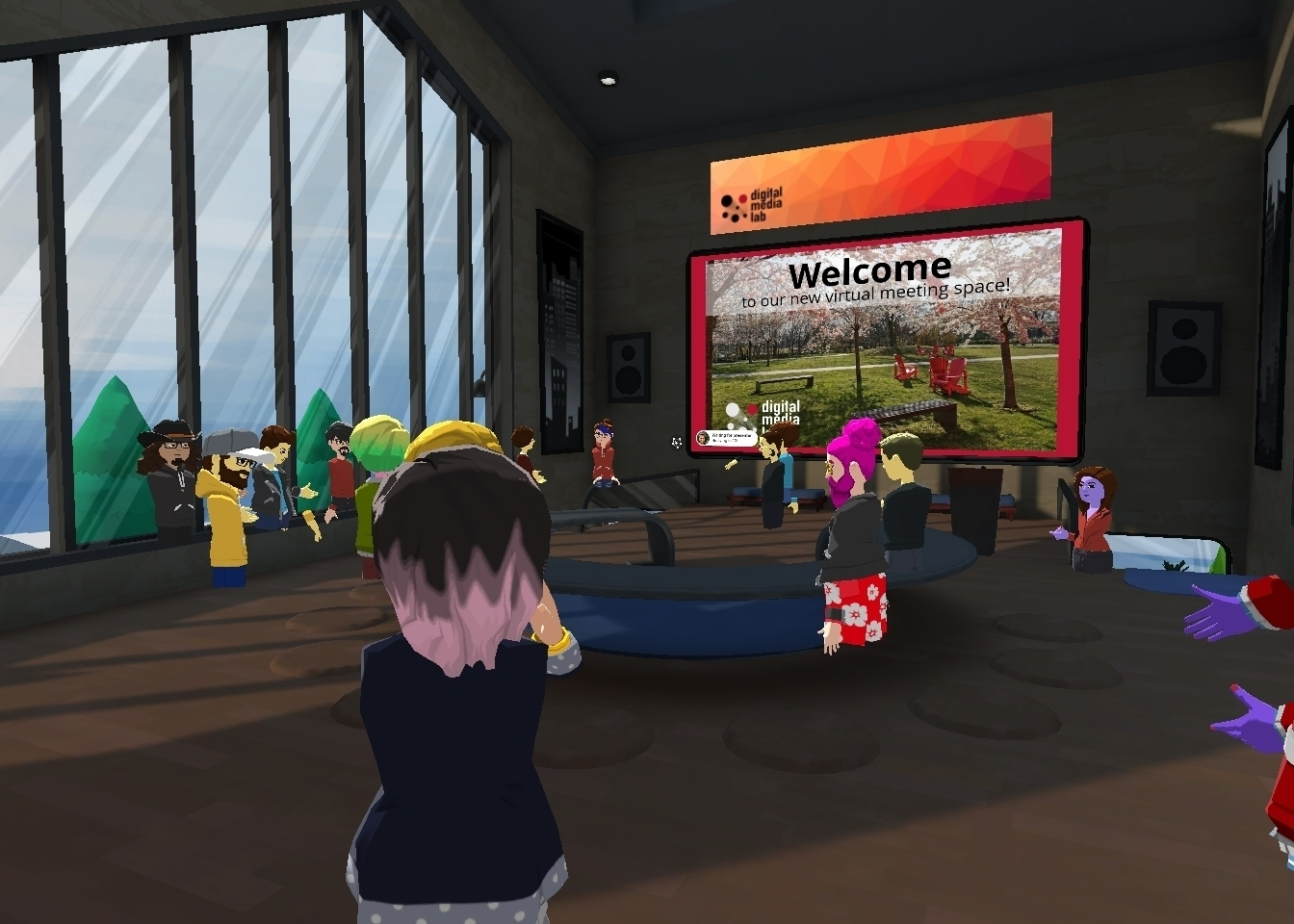}
\end{center}
\caption{This shows the setup of the \textit{AltspaceVR} platform. It was an indoor room with large windows and a presentation screen.}\label{fig:altspace}
\end{figure}

\begin{figure}[!ht]
\begin{center}
\includegraphics[width=13cm]{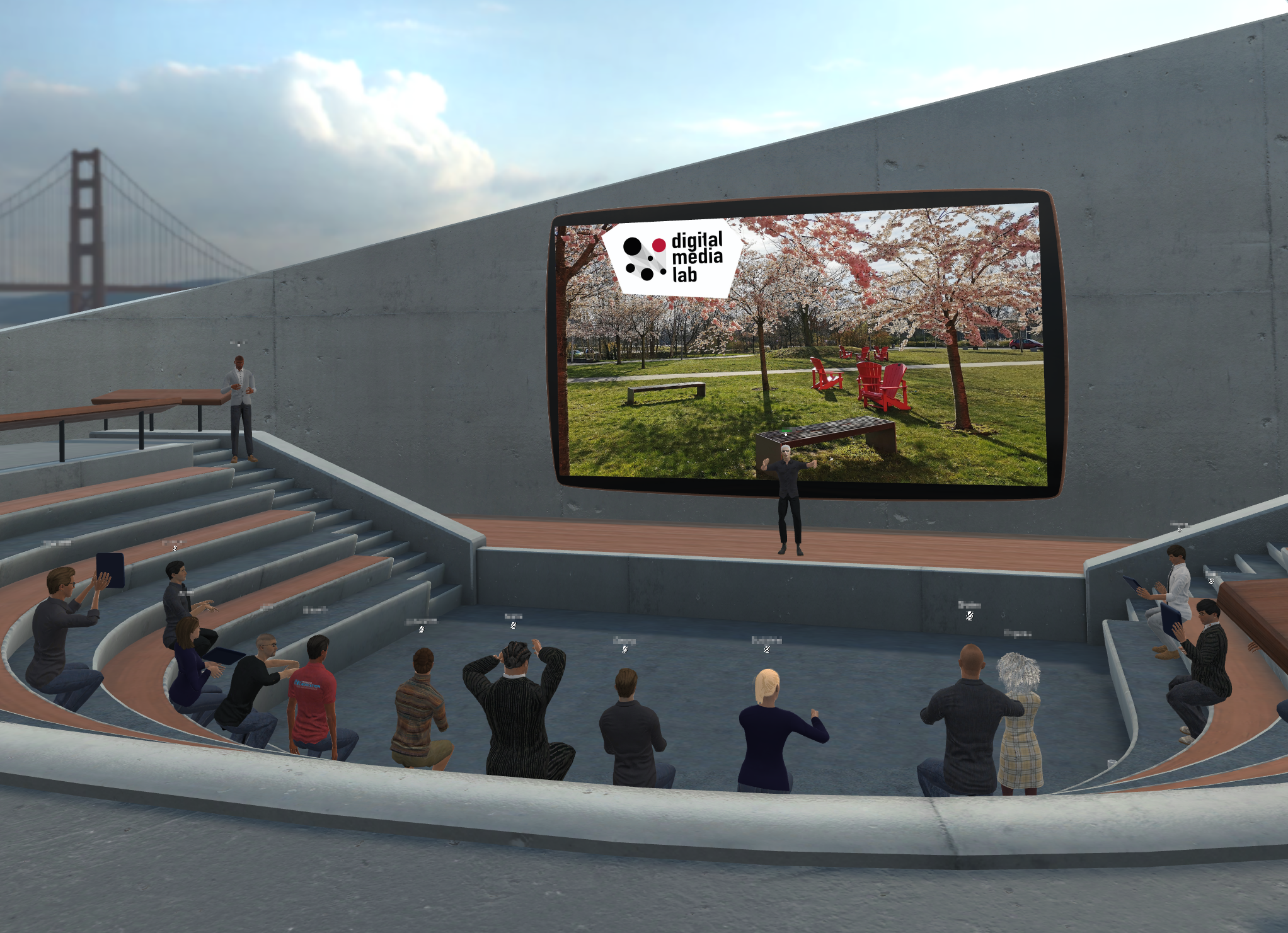}
\end{center}
\caption{This shows the setup of the \textit{Engage} platform. It was an outdoor scenario similar to an amphitheater with a presentation screen.}\label{fig:engage}
\end{figure}

\begin{figure}[!ht]
\begin{center}
\includegraphics[width=13cm]{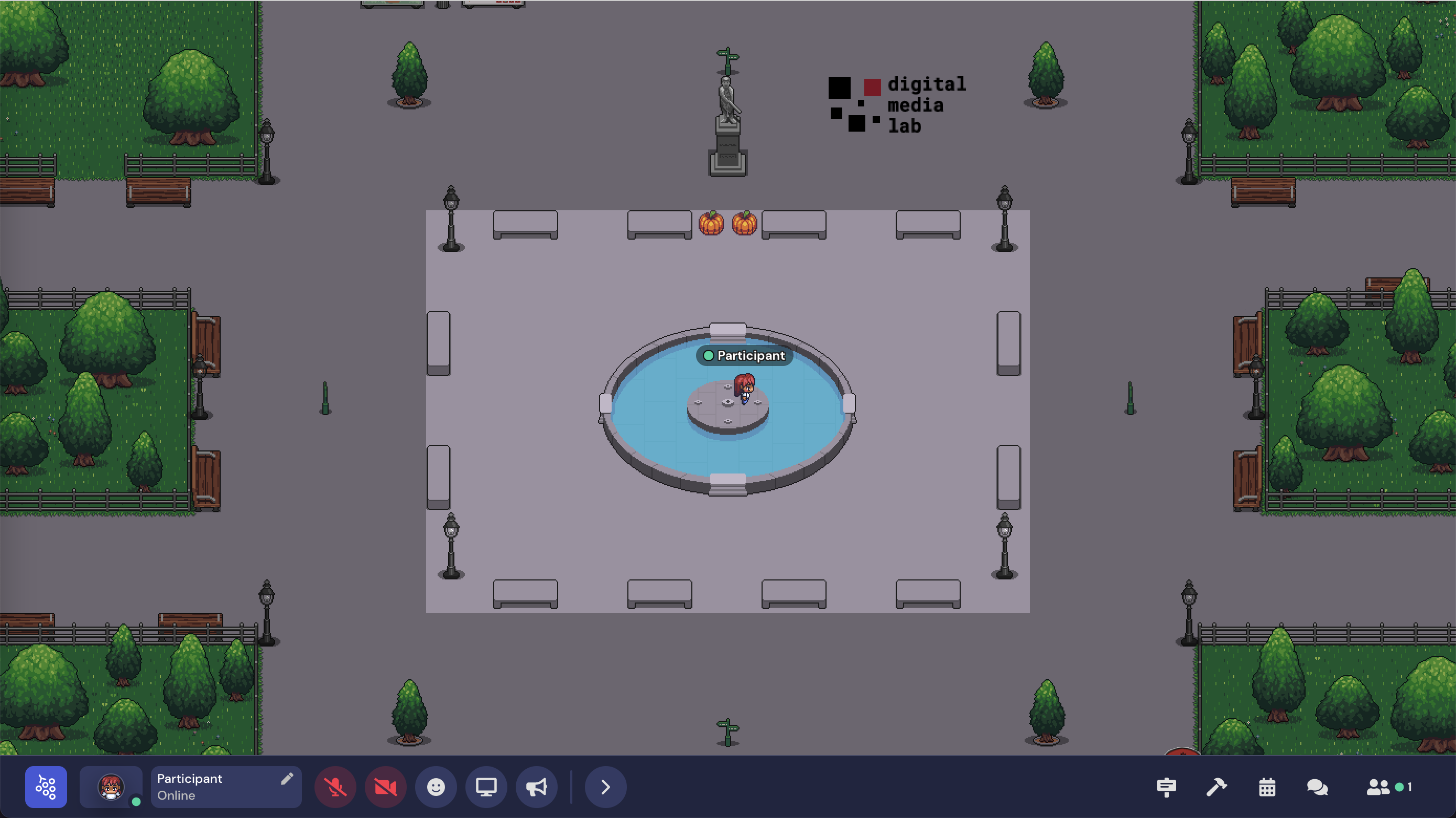}
\end{center}
\caption{This shows the setup of the \textit{Gather Town} platform. The scenery was an outdoor place in a park without a dedicated presentation screen. The lighter grey space identifies one coherent meeting area.}\label{fig:gather}
\end{figure}

To gain insights into the suitability of openly available social VR services for team meetings in the wild, we solely used established and sophisticated platforms that allow integrating presentations and giving talks. %and offer high immersion. 
The most critical factor for the software selection was the accessibility from all devices used by the attendees, as they were free to choose which device they preferred to access each meeting. 
Throughout the case study, we tested the VR platforms \textit{AltspaceVR}\footnote{\url{https://altvr.com}} (see \autoref{fig:altspace}, meetings 2--6) and \textit{Engage}\footnote{\url{https://engagevr.io}} (see \autoref{fig:engage}, meetings 7 \& 8), the videoconference platforms \textit{Zoom}\footnote{\url{https://zoom.us}} (see \autoref{fig:teaser} (1), meetings 9, 10 \& 12) and \textit{StarLeaf}\footnote{\url{https://starleaf.com}} (meeting 1), as well as \textit{Gather Town}\footnote{\url{https://gather.town}} (see \autoref{fig:gather}, meeting 11), a hybrid of a virtual 2D environment and video calls. We provide additional screenshots from the meetings and a video figure in the supplemental materials.

The virtual setup of AltspaceVR was based on a template provided by AltspaceVR. We used a template from the category ``Talk Show'' because it had a big screen for presentations and consisted of a room with large windows. This layout was inspired by the real setting the meetings were held in pre-Covid. We modified the virtual space with the group's logo as visible in \autoref{fig:altspace}. The table also resembled the U-shaped setup of our physical meeting room and gave the virtual room some structure. Participants could customize their avatar representation, mute themselves, react with emoticons, and move around the space. There was a room admin who could enable the megaphone feature for single persons to make them audible to the group in the whole room.

Similar to the setup of AltspaceVR, the meeting space of Engage was based on a template provided by the platform. The room design resembled an ancient outdoor forum with a U-shaped area to sit down to watch a stage. The stage featured a large presentation screen in the background. The meeting area also featured some educational exhibits and was placed on an island, viewing the Golden Gate Bridge in the distance. In contrast to the AltspaceVR settings, the people could actually sit down and were evenly spaced in the VE, creating the impression of a seated audience for the person on stage. 

We relied on a pre-existing template for the Gather Town meeting space again. We chose a park setting (see \autoref{fig:gather}) featuring areas for socializing with games. We modified the template so that the area around the fountain was one meeting space -- which is colored in light grey in \autoref{fig:gather}, where everyone inside could see the video of everyone else in this area (cf. \autoref{fig:teaser} (4)). Additionally, we added the two pumpkins between the upper benches to mark two tiles with a broadcast feature. The participants standing on these tiles were audible in the whole environment, similar to the megaphone feature of AltspaceVR. Outside of dedicated meeting areas, the visibility and audibility depended on the proximity of the participants, allowing the creation of small spontaneous groups.

\subsection{Data Analysis} 
The questionnaires collected quantitative data (7-point Likert scales) and qualitative (answers to open questions). The quantitative data were analyzed using factor analysis. As the questionnaires included various attributes that are not independent of each other, an exploratory factor analysis (EFA) with Varimax rotation on all questionnaire items was used to reveal hidden \emph{factors} that group multiple related questionnaire items. This data-driven process allowed us to identify groups of questions that influenced the users' ratings of their meeting experiences.
The qualitative data of the questionnaires were analyzed using thematic analysis by identifying similarities and inductively creating themes. A team of four researchers individually coded three responses, and after that, they discussed and iterated the resulting codes until they agreed on a common code system. This system was used to code all qualitative responses once. We received 200 responses. 155 of them contained qualitative data in addition to the questionnaire items. After the additions to the code system from the first cycle were discussed, a different researcher coded all responses a second time. This process resulted in 1,228 codings using 107 codes.

\section{Results \& Lessons Learned} 

In the following, we provide our analysis of the quantitative data collected in subsection~\ref{sec:ResultsRatings} as well as insights on the five themes derived from the qualitative data in subsections~\ref{sec:ResultsSpatial} to \ref{sec:ResultsUserNeeds}. The questionnaire responses to the Likert scale questions are evaluated with a focus on the ratings of the meeting experience. To provide authentic impressions and derive and discuss dominating themes that explain and contextualize the quantitative results, we present the qualitative insights in greater depth in the following. We report on the experiences and statements of the participants on the five main themes that the thematic analysis revealed: Spatial Aspects, Meeting Atmosphere, Expression of Emotions, Meeting Productivity, and User Needs. Each subsection concludes with a Lessons Learned paragraph discussing central insights and linking them to related work.

\subsection{Ratings of the Meeting Experience}
\label{sec:ResultsRatings}
As part of our post-meeting questionnaire, we obtained feedback on 19 Likert-scaled question items, shown in \autoref{table:questions}. In the following, we present the results of our quantitative data analysis of the users' ratings on those items.

\begin{table}
\centering
\caption{Overview of the most important questionnaire items. Questions from standardized instruments are denoted with \textit{SPQ} for the Social Presence Questionnaire by \cite{Nowak2003.SocialPresence} or \textit{UBS} for the User Burden Scale by \cite{suh2016.UserBurdenScale}. The constellations of all five questionnaires is detailed in the supplemental material.}
\label{table:questions}
\begin{tabularx}{\linewidth}{lX}
\toprule
                      Attribute &                                           Question \\
\midrule
    face\_to\_face\_meeting \textsuperscript{SPQ} & To what extent was this like a face-to-face meeting? \\
  same\_room\_with\_partner \textsuperscript{SPQ} & To what extent was this like you were in the same room with your partner? \\
         partner\_realism \textsuperscript{SPQ} &      To what extent did your partners seem “real”? \\
      choose\_to\_persuade \textsuperscript{SPQ} & How likely is it that you would choose to use this system of interaction for a meeting in which you wanted to persuade others of something? \\
      get\_to\_know\_extent \textsuperscript{SPQ} & To what extent did you feel you could get to know someone that you met only through this system? \\
     meeting\_involvement &                I felt involved in today's meeting. \\
    active\_participation &        I participated actively in today's meeting. \\
            felt\_noticed &          I felt noticed by the other participants. \\
   felt\_group\_membership &               I felt like being part of the group. \\
    meeting\_productivity &                 We had a productive meeting today. \\
  worry\_information\_leak \textsuperscript{UBS} & $<$The meeting platform$>$ accesses and uses the device's microphone and camera.  I'm worried about what information is being passed on by it.\\
     privacy\_trustworthy \textsuperscript{UBS} & $<$The meeting platform$>$'s policy about privacy is not trustworthy. \\
     privacy\_requirement \textsuperscript{UBS} & $<$The meeting platform$>$ requires me to do a lot to maintain my privacy within it. \\
      moderation\_quality & The moderation and guidance by the moderator in today's meeting were ... \\
        agenda\_structure & The agenda and structure of today's meeting were... \\
     vr\_meetings\_replace & I think the VR meetings will replace the video conferences in our lab. \\
  technical\_difficulties &                We had many technical difficulties. \\
        usage\_confidence &           I felt confident about using the system. \\
total\_meeting\_experience &    Overall, how was your meeting experience today? \\
\bottomrule
\end{tabularx}
\end{table}

\begin{figure}
    \centering
    \includegraphics{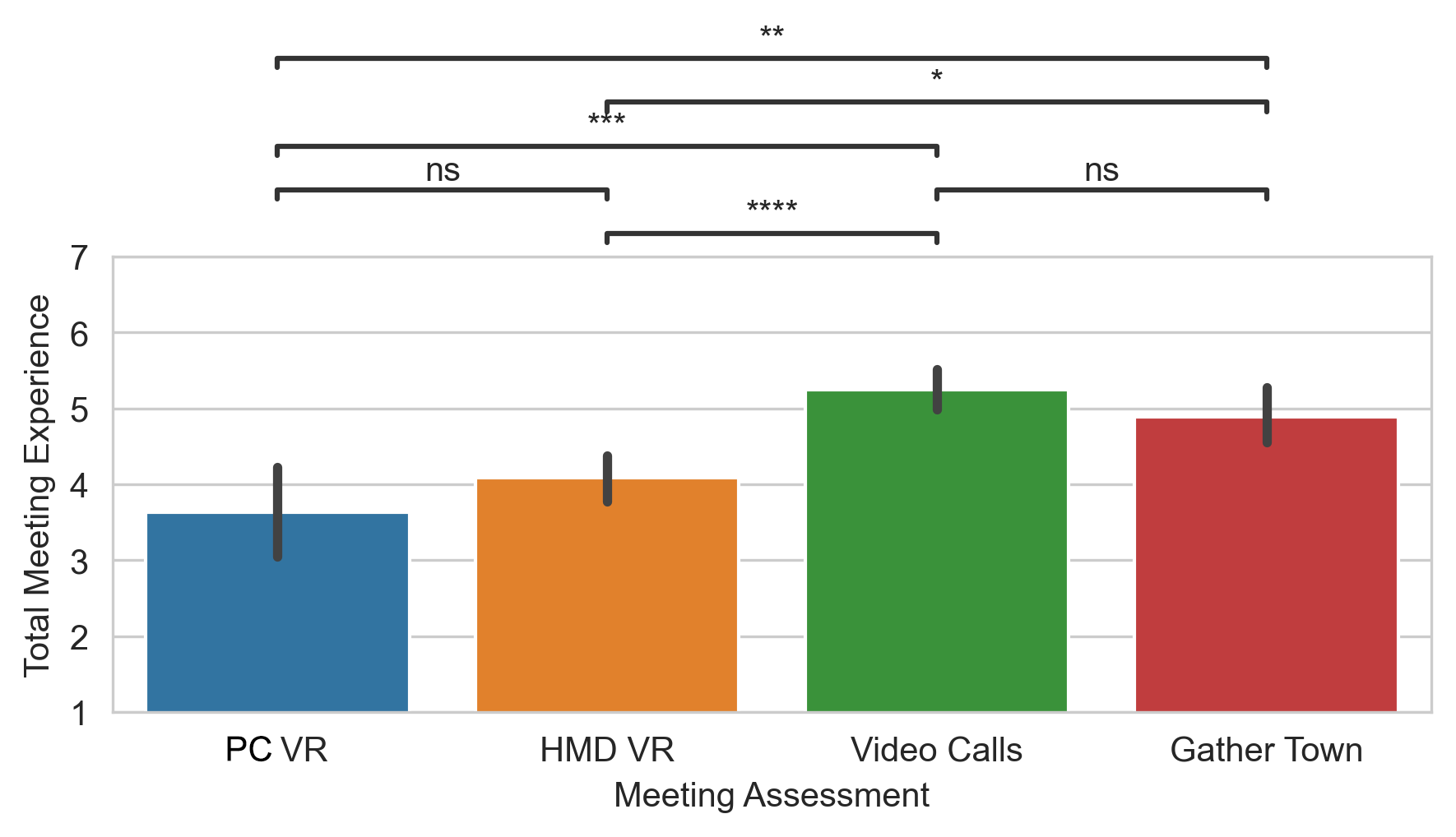}
    \caption{Barplot of the total meeting experience assessed in meetings M2--M12 together with the results of Bonferroni-corrected paired t-tests. }
    \label{fig:total_meeting_exp}
\end{figure}

% Total Meeting Experience
Starting from meeting M2, we asked the attendees about their overall meeting experiences on a 7-point Likert scale (poor $1$ to $7$ great). Figure~\ref{fig:total_meeting_exp} shows a barplot with the results of the total meeting experience. 
With a one-way ANOVA, we found a significant effect of the meeting modality on the total meeting experience ($F(3,178){=}11.041, p{<}0.01, \eta_{p}^{2}{=}0.157$).
Bonferroni-corrected post-hoc tests confirmed that the total meeting experience in the video call meetings was significantly higher than in the VR meetings for both VR options, VR via desktop PC ($t(30.454){=}{-4.876}, p{<}0.01, d{=}{-1.429}$) and VR using an HMD $t(137.635){=}{-5.340}, p{<}0.01, d{=}{-0.821}$). For Gather Town, we also observed an improved meeting experience for PC VR ($t(35.009){=}{-3.509}, p{<}0.01, d{=}{-1.062}$), and HMD VR ($t(45.499){=}{-3.124}, p{=}0.019, d{=}{-0.537}$), and no difference for the video call platforms ($t(35.186){=}{-1.461}, p{=}0.917, d{=}{-0.372}$).
These results indicate an advantage of video calls and the hybrid platform Gather Town compared to the VR conditions. 
Nevertheless, we did not observe any difference between the VR options $t(35.156){=}{-1.321}, p{=}1.00, d{=}{-0.290}$) and, therefore, we did not differentiate between these two VR variants in the following quantitative analysis.

% Correlation with Pre-Experience
Since the attendees had a very diverse pre-experience with VR technology, we investigated whether the pre-experience, as assessed in meeting M1, correlates with the meeting experience in the first VR meeting M2. Using a Spearman rank correlation, we found that the two variables were moderately correlated (Spearman's $r(13){=}0.514$, $p{=}0.072$).

However, the total meeting experience rating did not provide a conclusive impression of the users' perception of the meetings. Only four participants attended all twelve meetings, which prevented a more detailed within-subject analysis in the following. Still, we wanted to investigate exploratively which attributes contribute to the user ratings to understand the observed differences.

\begin{figure}
    \centering
        \includegraphics[width=13cm]{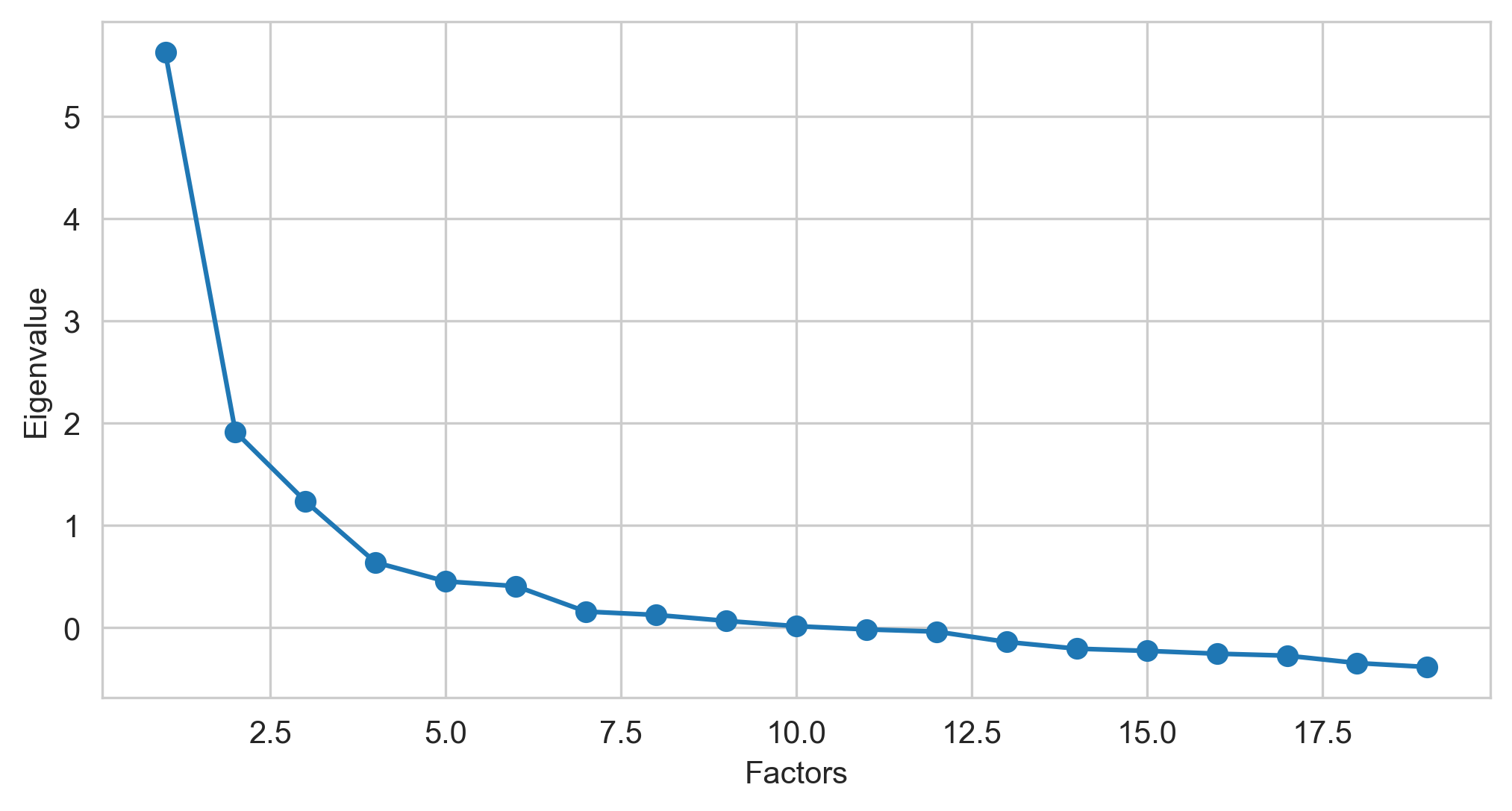}
        \caption{Scree plot of variance explained by each factor, resulting from the factor analysis on questionnaire items.}
        \label{fig:efa:scree}
\end{figure}

\begin{figure}
    \centering
        \includegraphics[width=7.5cm]{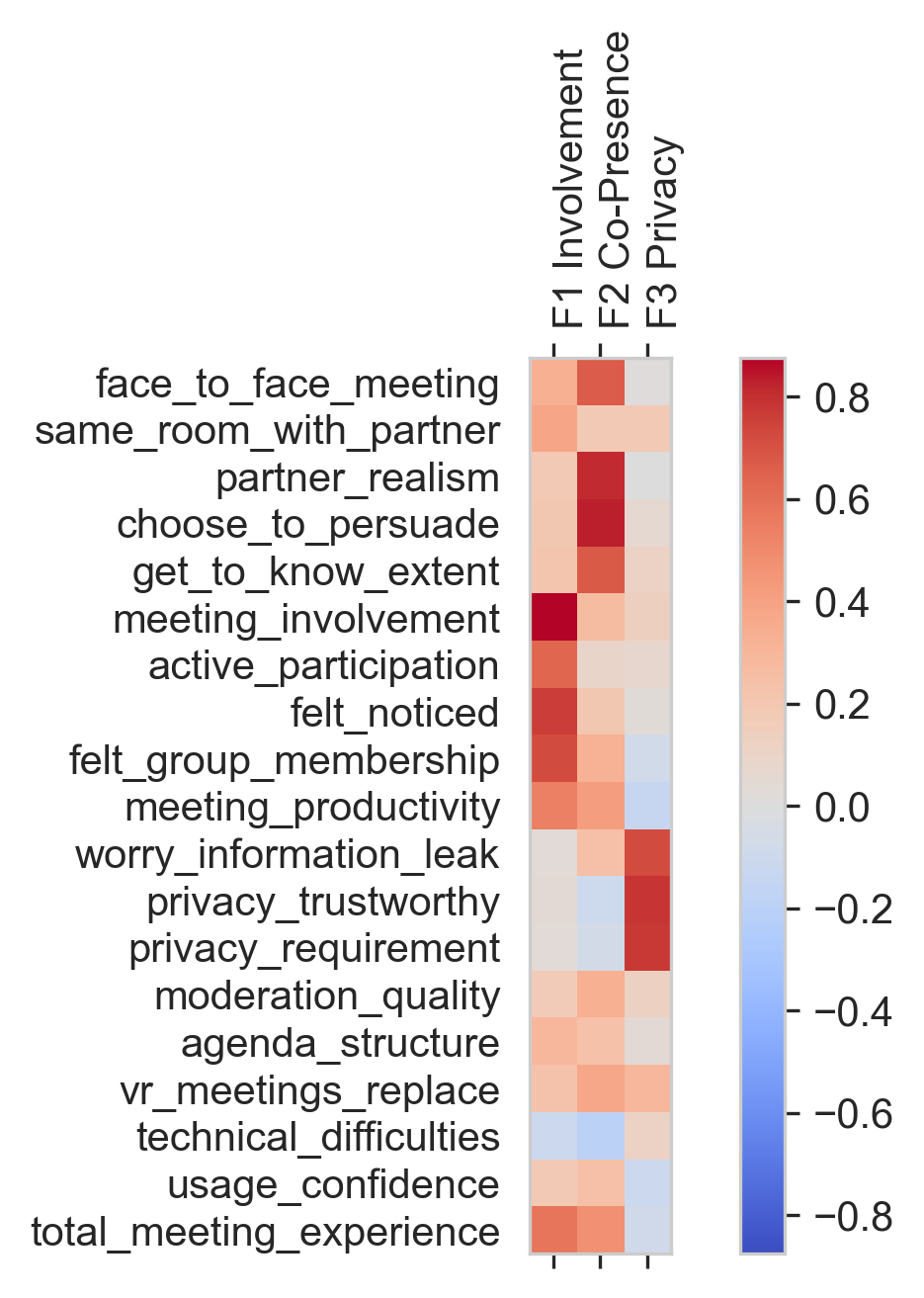}
        \caption{Loadings of all attributes (larger absolute values represented in darker colors) for the three factors (F1--F3) in the exploratory factor analysis.}
        \label{fig:efa}
\end{figure}

\textbf{Exploratory Factor Analysis. }
To explore the structure of user ratings and identify possible factors in the multiple questionnaire items of Table~\ref{table:questions}, we conducted an exploratory factor analysis on the additional feedback obtained through the Likert scale questionnaire items. Figure~\ref{fig:efa:scree} presents a scree plot of the variance associated with each factor, representing the explained variance for each resulting factor. This helps us to obtain the number of factors. Factors with high levels of explained variance are important, interpretable contributors to the overall model, whereas later factors explain less variance. The scree analysis revealed three factors with an Eigenvalue above~$1$. Continuing with these three factors, we calculated the factor loadings. Figure~\ref{fig:efa} shows the loadings of all question items for the three factors. The loading of an attribute for a factor can be interpreted as the correlation between the question item and the factor. Therefore, the higher the absolute loading value, the stronger that attribute is tied to that factor and predicted by it. 
On the other hand, a loading of close to 0 indicates the absence of a (linear) relationship between the attribute and the factor. 

This allowed us to find interpretations of these factors. Factor F1 loads highest on the questions related to \textit{Involvement}, F2 corresponds to \textit{Co-Presence} items, and F3 to items around \textit{Privacy}. 
Subsequently, we compare VR meetings, video calls, and Gather Town with each other according to the three factors. % as shown in \autoref{fig:factorplots}.  
With a one-way ANOVA, we found a significant effect of the meeting modality on factor \textit{F1 Involvement} ($F(2,156){=}4.263, p{=}0.016, \eta_{p}^{2}{=}0.052$).
Bonferroni-corrected post-hoc tests confirmed a difference between VR and video calls with $t(126.17){=}{-3.195}, p{<}0.01, d{=}{-0.515}$ with higher ratings for video calls.
Another one-way ANOVA revealed a significant effect of the meeting modality on factor \textit{F2 Co-Presence} as well ($F(2,156){=}26.934, p{<}0.01,\eta_{p}^{2}{=}0.257$). Post-hoc tests confirmed a difference between VR and video calls with $t(137.940){=}{-8.228}, p{<}0.01, d{=}{-1.269}$, and between Gather Town and video calls with $t(16.149){=}{-3.775}, p{<}0.01, d{=}{-1.446}$ each with better ratings for video calls.
We could not find an effect for \textit{F3 Privacy} ($F(2,156){=}0.309, p{=}0.735,\eta_{p}^{2}{=}0.004$), indicating conclusive experiences on that factor independent of the meeting modality. These results demonstrate the advantages of video calls in contrast to the other platforms. 

\textbf{Lessons learned. } Concerning quantitative data, we could observe that video meetings outperformed VR meetings in terms of the total meeting experience. Furthermore, for two of the three factors revealed with the exploratory factor analysis, i.e., \textit{F1 Involvement} and \textit{F2 Co-presence}, we also found that video conferencing was ranked higher than VR meetings. Users tended to feel more involved and have higher co-presence with their colleagues in video calls than in VR meetings. With the immersive nature of VR and its capability to substitute the user's physical environment, it is surprising that the attendees had a stronger impression of others being present on video calls than in VR. 
To obtain a deeper understanding of the reasons for this assessment, we provide detailed insights on the experiences along the five themes that evolved from our qualitative analysis in the following subsections.

\subsection{Spatial Aspects}
\label{sec:ResultsSpatial}
An essential difference between video calls and virtual 3D environments is the spatial component that allows movement in space and varying proximity to others. It influences the group dynamics and the audio considerations.

\begin{figure}[!ht]
\begin{center}
\includegraphics[width=13cm]{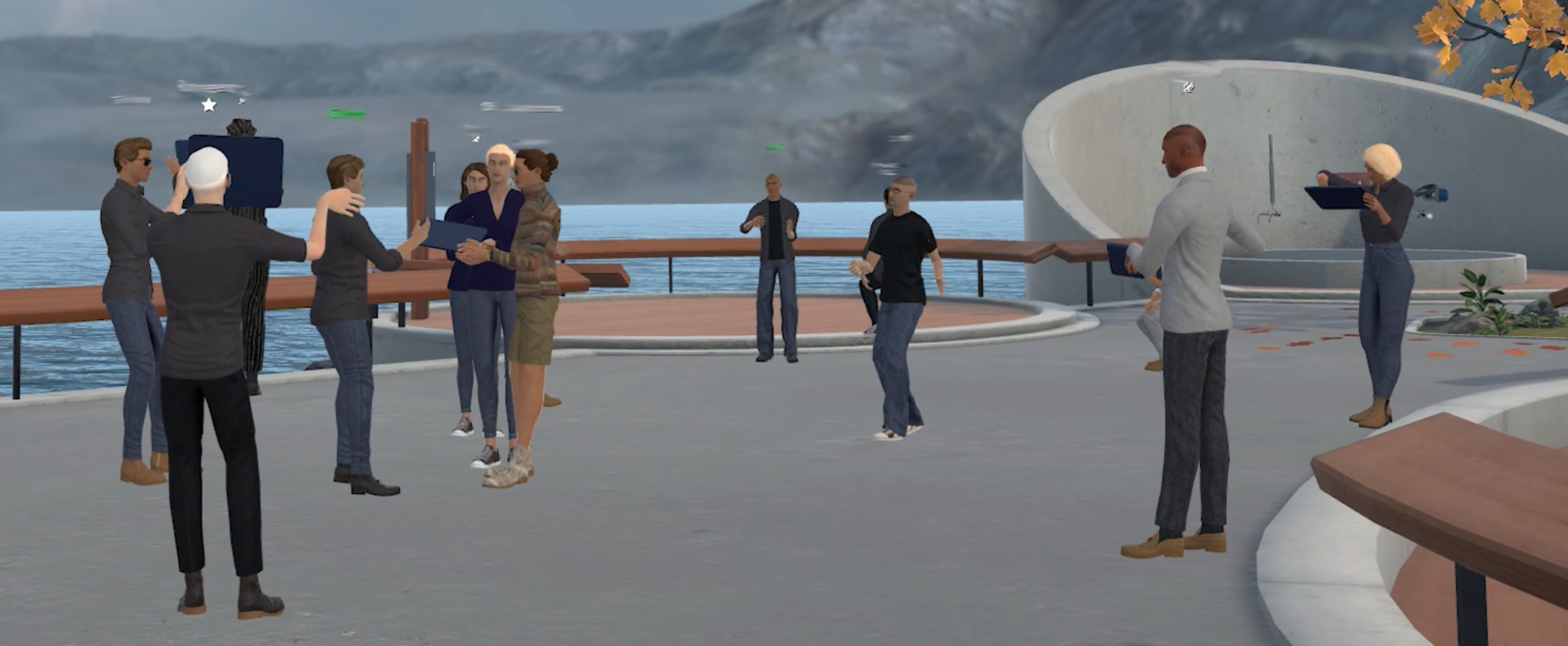}
\end{center}
\caption{A situation in Engage before the meeting where the participants grouped (left side) and one person walked up to this group (right side, black shirt).}\label{fig:groups}
\end{figure}

\textbf{Movement. }
Before the study, P08 expected \says{more fun and more movement during the meetings} in VR. Indeed, the possibility of moving around in the VE led to more vivid and \say{a lot more dynamic}{18} meetings compared to both video calls and physical meetings. Some attendees preferred that everyone be seated for the meeting to be more \say{orderly}{28}. 
Other participants appreciated \say{being able to move around and arrange in the room}{12} and to \say{walk in front of the screen and use [their] body}{18} during presentations. %This was appreciated during presentations as they .%, but the presentation aspect is further discussed in its own (sub)theme.% 
This was sometimes perceived as troublesome during presentations because avatars of other attendees were blocking the view onto the slides or the speaker: \say{All the time there was someone in my way}{09}. However, for presenting, the participants described advantages: \say{I also liked the more true-to-life presentation environment with the projector screen and with the ability to see the audience spread out in a semicircle instead of a gallery view like on Zoom}{07}. % Another presenter appreciated that \say{I could walk in front of the screen and use my body}{18}. 
Many attendees highlighted the value of meaningful movements such as waving, gaze direction, and pointing as indicators, e.g., to initiate conversations or for turn-taking, making it \say{more personal}{16}. That \say{you could walk up to someone to talk to them}{17} was perceived as more effortless and more natural (see \autoref{fig:groups}). %Further, hugging each other was a meaningful movement: \say{When I met Linus, we first shook hands. Then we decided that a hug would be more appropriate, and it was a really nice moment}{18}.%Also waving was described as a way to initiate conversations in a VE which \says{feels more personal} according to P16. 
Another use of the spatial component was to form queues, especially after the meeting when several attendees wanted to speak to the same person. 

\textbf{Proxemics. }
Similarly to movements, proxemics was used with purpose or meaning. \says{The physical arrangement also plays a role -- who sits next to whom}, mattered for P12, or when everyone in a circle is supposed to say something in turn. At the same time, the spatial relations did not always translate well to the VE: \say{In physical meetings, it is nice that you actually sit next to others. In VR, it still feels a bit odd (am I too close? too far away?) -- it is also a bit more difficult to exactly place yourself the way you would like}{12}. Other attendees also reported that they \say{felt awkward}{05} or \say{felt bothered because the others came too close}{16}.

\textbf{Group Dynamics. }
The added spatial component of VEs allowed to form and vary group constellations and sizes, which was a crucial benefit to a majority of the participants:
%One of the major advantages of flexible spatial arrangements was the possibility to form and vary groups: 
\says{The VR environment makes it much easier to converse in small groups and easily switch between talking to different people}, explained P07.  
P06 pointed out that especially \says{before and after the meeting [interactions] were a lot easier and closer to in-person interactions than they would be in a Zoom meeting.} %This advantage was crucial to a majority of our participants. 
The participants described a large variety of interaction types, including 1-to-all as in presentations, 1-on-1, small groups, large groups, the whole group, and dynamic transitions between the constellations. 
Other than in Zoom, small groups could split up and follow up on meeting topics in parallel, which is \say{barely possible [...] next to the official round}{12} without blocking the meeting room for all others in a video call. Therefore, VR meetings were described as more dynamic, allowing more attendees to say something and contribute. On the other hand, Zoom meetings were characterized as \say{static}{14}.

VEs also allow the users to position themselves in a third dimension: elevation. This affords extended group arrangements and was primarily used for presentations and other 1-to-all communication. One participant experienced the group manager hovering mid-air above the rest of the team as a status imbalance: \say{It was beneficial that we could all see him. But it also gave the impression that he was `superior' and talking down on us}{18}.
% His spatial position translated into a status position in my head -- hovering like Jesus

\textbf{Spatial Audio. }
The systems based on a VE are designed so that a conversation is only audible to users standing close by. The further away a listener is, the quieter the sound becomes. 
While this enabled several small groups to have private conversations simultaneously in the same room, it was perceived as cumbersome for the plenum when everybody should hear everybody else by default: \say{In case of a presentation or group discussion, sometimes you don't understand others and we used the `megaphone' [to amplify voices in AltspaceVR] all the time}{05}. %There, it was seen problematic by P05: .
Furthermore, attendees reported that, also in VR, \say{the main questions you hear are: `am I muted?' or `can you hear me?'. The spatial audio is completely useless.}{05}. Manually equipping speakers with the `megaphone' as an amplifier was referred to as \say{a burden}{01}, not only by the participants who facilitated the meeting but also by observers. 
%``pretty much everytime someone talks he/she needs the megaphone modifier, which is an additional administrative hassle (to activate and deactivate)'' \#666667 -- 16.9.
The thresholds for the spatial audio were often experienced as inadequate resulting in unnatural behavior: \say{Especially because of the spatial audio, people gathered closely in clusters so they can hear each other better. In reality, it wouldn't be that condensed, and also not so hard to hear each other}{18}. Instead, P05 suggested \says{to have regular audio [for the plenum] and for private conversations sound bubbles.}

\textbf{Lessons Learned.} The VEs allowed the attendees to gather in a virtual workspace. Videoconference platforms, such as Zoom or Skype, recently introduced features to create a similar impression by visually cropping attendees and stitching them together in a shared room\footnote{\url{https://blog.zoom.us/introducing-zoom-immersive-view} \\ \url{https://www.skype.com/en/blogs/2021-05-together-mode}}. However, this lacks intradiegetic possibilities to utilize this added spatial component, for example, the eye gaze in turn-taking.
When designing a VR meeting platform, we consider it essential to foster dynamic interpersonal interactions by supporting meaningful movements, such as approaching others, forming groups and queues, waving, pointing, or nodding. Non-verbal cues strongly impact social interactions in VR but have not yet tapped the rich potential familiar from offline behavior \citep{tanenbaum2020.ExpressiveNonverbalCommunicationinSVR, maloney_talking_2020}.
In line with related research, our participants reported spatial group arrangements and proxemics similar to in-person meetings. As \cite{Williamson2021.ProxemicsSocialVRWorkshop} point out, elevation provides additional opportunities for an audience in VEs to arrange for better visibility. While this can be beneficial, we advise caution when activating flying features, as they might diminish professionalism or affect the perception of social status differences.

For flexible and realistic group dynamics, purposeful and configurable spatial audio with carefully chosen volume reduction parameters is required, as well as an option to hear everyone in a plenum. We observed similar behavior of our participants as in a study by \cite{williamson_ProxemicsAudio_2022}, where users seemed to move close together to hear each other best despite loud background conversations. As our participants proposed, allowing users to create a private sound bubble with visible boundaries to shield their group might be helpful.
Already in 1999, \citeauthor{Stahl1999.ancientVRMeetings} reported in a similar case study that the uncertainty of being audible was one of two critical challenges for VR meetings. Even today, we still observe the need for more explicit indicators of one's comprehensibility within the room.

\subsection{Meeting Atmosphere} 
We further observed that the spatial component strongly influenced the meeting atmosphere. In the questionnaire before the first meeting in VR, P18 shared that the \says{feeling of working in the same place is missing} during the long home office directive. The potential of immersively feeling \say{like being somewhere else}{12} for the meetings promised relief to some participants.

\textbf{Co-Presence and Interpersonal Interactions. }
Numerous attendees expressed excitement about \say{being in a room together}{06} and \say{chatting `face-to-face'}{12}. 
For P18, in the VE, it felt more like \says{coming together or gathering with the whole team} in a shared virtual office space than in video calls. P23 and P09 were reminded of physical meetings, which included that \say{interactions felt more like a real meeting}{09}. The effect was perceived to be stronger when accessing the VE with an HMD: \say{I felt more immersed and part of the group}{09}. P25 appreciated how \says{one has a better awareness for the group, such as closeness or distance, attention or distraction, etc., which is missing on Zoom.}
While in VR, participants appreciated seeing everyone around them at a glance, the videoconference tools were repeatedly criticized for not displaying all attendees in large meetings: \say{I frequently do not register that they are attending the meeting}{12}. This is especially problematic when presenting as it \say{gives a wrong impression about the number of participants}{11}. 

P18 appreciated being \says{surrounded by my colleagues, but without faces, it was not so personal. Like a coat of secrecy around everyone.} With the cartoony avatars on AltspaceVR, to P23, it \says{felt like speaking with very intelligent and responsive game characters}. This dilemma was especially noticeable for guests. Several team members expressed the wish to see a guest's face \say{to get an idea of how [they are] as a real person}{28}. On the other hand, the avatar representation allowed guests to \textit{``visit''} the meeting as equal to the other members.%while in a physical meeting or on video call they would be \say{just calling}{18}.

\textbf{Professional or Playful. }
The initial atmosphere of the meeting was established mainly by the visual appearance of the interface, the avatars, and the environment. %While Zoom and StarLeaf show neutral and business-oriented interfaces, Gather Town and AltspaceVR have a more playful and fun graphic design with comic-like avatars: 
The setting of AltspaceVR was perceived as \say{sometimes too cartoonish to be taken seriously. [...] more like in-game conversations}{23}. % Engage strives for a professional appearance.
On the other hand, Engage was perceived as more formal and \say{much more orderly than in Altspace because we were seated}{28}. 
%Also the visual style resembled in-person encounters: \say{As the avatar is more real, it looks more like a physical meeting}{13}. 
Both the professional and the playful styles have received positive and negative feedback from the participants. Some participants liked the professional atmosphere that Zoom or Engage created because it is similar to physical meetings and makes it easier to talk about serious topics. But it was also perceived as more boring and formal, and therefore some rejected the approach of Engage: \say{I really hated Engage. The uncanny avatars, technical difficulties, and overall `seriousness' of the application didn't work for me}{09}.  Due to the less formal setting in AltspaceVR, \say{even the student was completely chilled -- usually on Starleaf they're more nervous and quiet}{18}. 
As Engage allows for adding elements to the scene, it was soon \say{filled with animals, objects, special effects, and sounds}{18}. This was described as \say{absurdly funny}{18} and \say{goofy}{02} but during the meeting also as \say{distracting}{17}, and \say{less effective}{01}.
Although the fun, interesting, and creative atmosphere in AltspaceVR, Engage and Gather Town can be appealing and support hedonic qualities of the system, it could be distracting from the subject of the meeting: \say{fun to use but not very helpful for serious work}{11}.

\begin{figure}[!ht]%{R}{0.6\textwidth}
\centering
  \includegraphics[width=0.6\columnwidth]{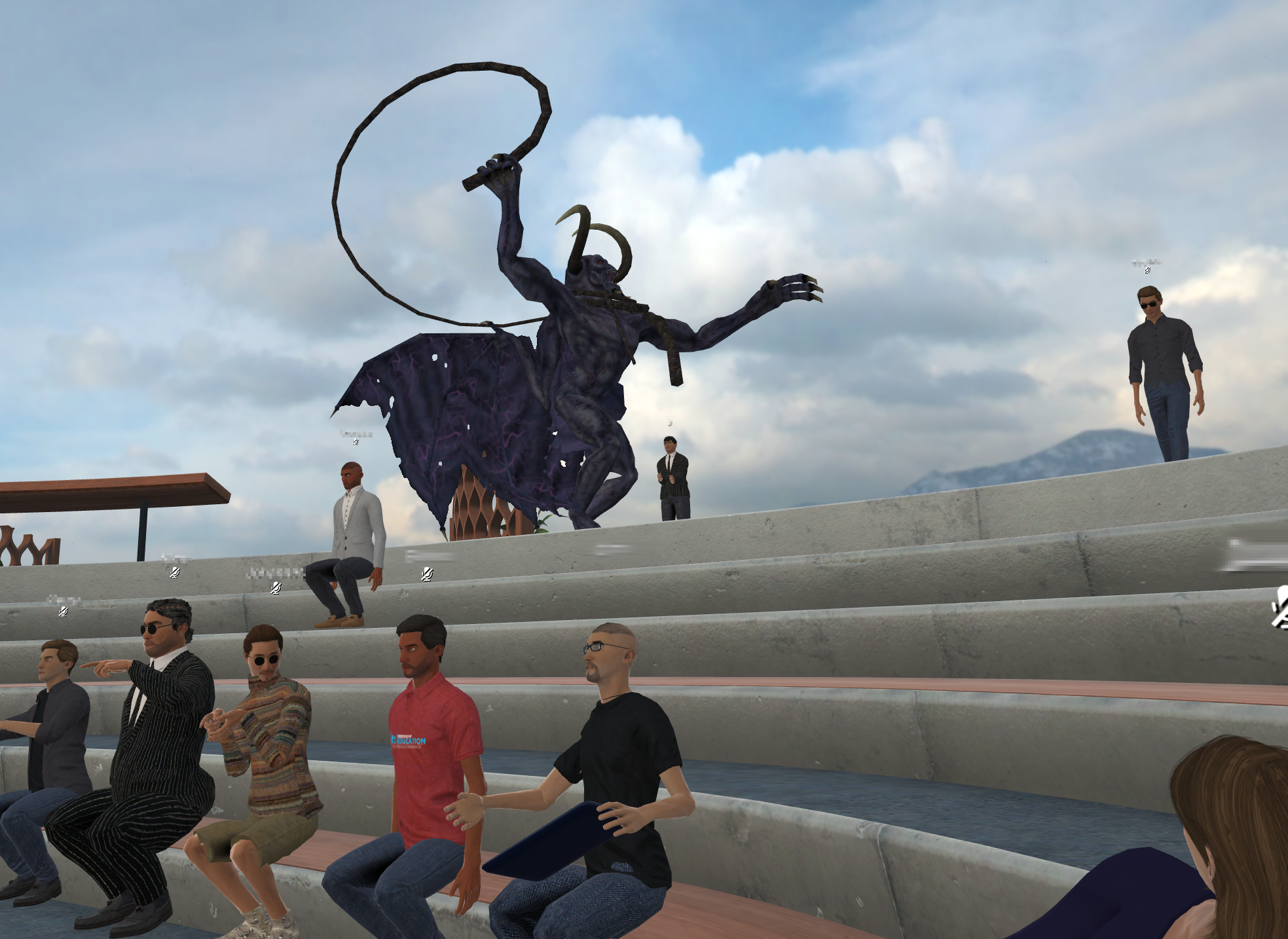}
  \caption{A user inserted a daemon to the scene whipping a colleague's avatar in the middle of a student's presentation.}
  \label{fig:daemon}
  %\Description{A screenshot from Engage showing a giant daemon with a whip.}
\end{figure} 

\textbf{Social Norms. }
Social interactions in VR depend on shared norms in a group as in reality. However, VEs enable behavior that is impossible in real life, such as walking through people, due to different physical or perceptual restrictions. The attendees shared experiences of how some social conventions translate well to virtual encounters: \say{I instinctively wanted to hug their avatar. When approaching, 1 meter away, I noticed that this feels wrong. In reality, we would never hug each other [...], so I expressed my compassion verbally only}{18}. 
More often, participants reported novel social situations that might require new conventions. We identified two types of violating social norms: unintended impropriety and intended provocations.

Among the unintentional norm violations are situations in which, for example, attendees \say{felt bothered because the others came to close}{16}, or had unnatural postures due to the controllers sitting on the desk, causing awkwardness.
One attendee described another uncomfortable situation due to technical challenges, in which he was concerned about having offended somebody with a joke as they did not react. It turned out that the other person was only staring at him quietly because they could not unmute to share the laughter.
Other disruptions, however, were intended by users, such as an \say{eccentric avatar}{18} that was seen as surprising at the professional occasion, somebody attending in a space suit, or users adding 3D models and effects to the scene. The most prominent example for this was \say{the monster which appeared in the middle of a student's presentation}{17}. The whipping daemon shown in Figure \ref{fig:daemon} caught the users' attention, distracted them from the presentation, and crashed the seriousness of the setting.

\textbf{Lessons Learned.} 
The users appreciated the impression that their colleagues were in one room with them, especially when using an HMD. \cite{Barreda-Angeles2022.PsychoEffectsSocialVRPresence} found social presence to be a good predictor of relatedness and enjoyment, which is in line with our participants' reports. The high loading values of the overall meeting experience for the factors \textit{F1 Involvement} and \textit{F2 Co-Presence} indicate a strong link between the predictors. Analysis of qualitative data corroborates this as especially those attendees, who appreciated the high presence of others and feeling noticed by others in VR, were most fond of those meetings. 
Still, attendees rated their Zoom experiences regarding \textit{F2 Co-Presence} higher than in VR. This finding is surprising considering the high immersion of VR and that many participants described the impression of being together in one room. However, the attendees of the investigated meetings are not strangers but colleagues, often for many years. Nevertheless, the avatars surrounding the user in VR seem unfamiliar and need a name tag to identify who it embodies. Without it, they could have been mistaken for intelligent game characters. In contrast, the identity of the conversation partners was unmistakable on video calls because of the visibility of familiar faces, which we discuss in the following subsection. 
Further, the ratings of the \textit{F1 Involvement} factor were higher for video calls. From an autoethnographic perspective, we suspect that the meeting format has been adapted too tentatively to the medium. To exploit the advantages of interpersonal interactions in VR, we emphasize that the meeting purpose must benefit from high social presence, e.g., with team building or socializing goals. Not every format should be directly replicated virtually, as previous research supports \citep{mcveigh-schultz_beyond_2021}.

The discomfort of the attendees with the uncanny degree of avatar realism appeared to be more pronounced than the increase in social presence found by \cite{zibrek2019.PhotorealismSocialPresence} as a consequence of photorealism in social VR. In line with related work, the effects of visual fidelity of our test platforms' avatars seem more complex and depend on other factors, such as coherence, setting, and avatar behavior. We suggest focusing more on expressive and versatile avatar designs than sheer photorealism.

We can assume that some of the observed social inadequacies originated from technical overload or accidents. For example, the sense of personal space is not transferred directly from in-person interactions to VEs and depends on various factors, such as using an HMD or a desktop PC \citep{williamson_ProxemicsAudio_2022}. A prior training and familiarization phase with the system~\cite{mcveigh-schultz_whats_2018} is advisable. This may also reduce disruptions from users exploring features that the platform affords. We urge interaction designers to create systems that prevent users from awkward appearance or behavior by accident, e.g. while taking off equipment or sitting down and standing up.
Further, until a common ``VRtiquette'' \citep{lahlou2021.BeyondBeingThere} is established over the years, the attendees should agree on shared rules and conventions that might be supported by regulations from the organizations -- in the same way as video conferencing and hybrid meetings benefit from this \citep{saatci_reconfiguring_2020}.

\subsection{Expression of Emotions}
The participants commented particularly much on the emotional aspects of interpersonal interactions and how replacing a video with an avatar changed their meeting experience. We must note that in the case of the team meetings investigated, most attendees typically turn on their cameras for the video call.

\textbf{Facial Expressions and Body Language. }
Seeing no facial expressions from the other attendees is one of the issues of VR meetings raised the most by our participants: \say{It is so odd to look at people and not see any reactions in their faces}{12}. Especially users who accessed the VE without an HMD and therefore lacked gesture and head tracking \says{looked a bit odd, less lively} to P12. The lack of faces was perceived as \say{disconcerting}{07}, \say{awkward}{06}, and \say{less direct, obtuse}{10} compared to video calls. Therefore, interactions on video calls were described as \say{more expressive}{14} -- to \say{see people's faces and emotions}{30} was important to connect for many. Several attendees have emphasized this concern throughout the experiment: \say{This is essential!}{12}. Above all, P02 missed the smiles and cats usually visible in the team's Zoom meetings. 
On the contrary, some attendees did not \say{need to see people's awkward or negative faces}{13} and were relieved to be able to hide behind their avatars. They highlighted the advantages of missing cameras: \say{It looked more like a physical meeting because I can see everyone at once and not just the face. I could see people moving their heads and hands in VR}{13}.
Several users addressed the expressiveness through avatar style and gestures: \say{I like the body language and the way people dress}{08}. One participant reflected on the impact that the different forms of encountering guests or strangers might have: \say{People present themselves differently. I even think you might get to know them differently in VR}{12}.
%It helps with all the social distancing going on}{30}. Other users found it difficult to recognize facial expressions on Zoom, though, with the large number of participants, \say{because you cannot see emotions that good on the small screens}{11}.

\begin{figure}[!ht]
\begin{center}
\includegraphics[width=13cm]{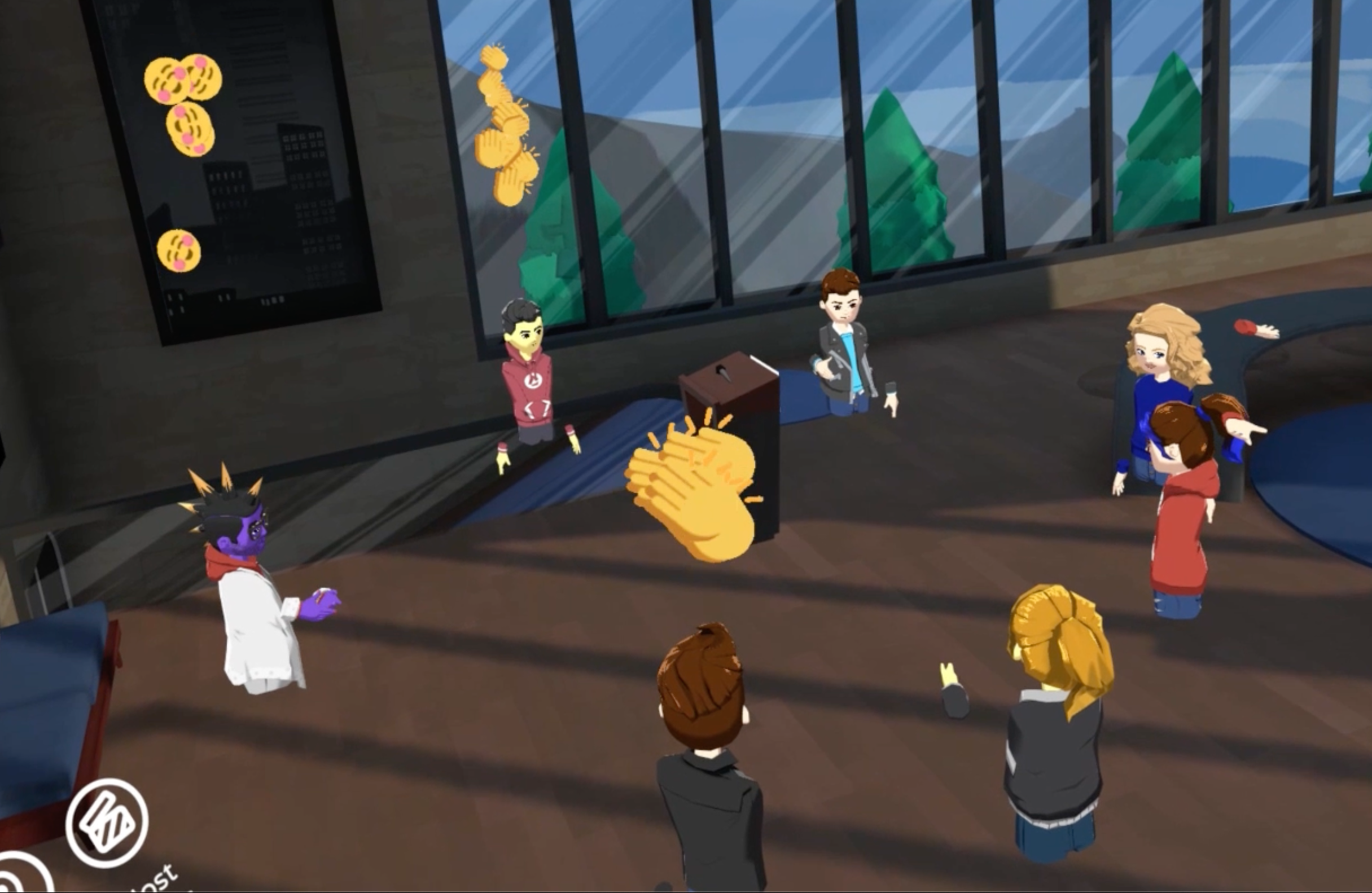}
\end{center}
\caption{This figure shows the use of emojis to express emotions in the AltspaceVR platform.}\label{fig:emojis}
\end{figure}

\textbf{Emojis. }
To compensate for the lack of facial expressions, AltspaceVR allows spawning emojis above the user's avatar to react visually (see \autoref{fig:emojis}). One attendee \say{expressed frustration with the –.– emoticons over her head}{18} while having audio issues and no other means to communicate.
A similar feature is also available on Zoom, despite other, more immediate ways to react: \say{When [the manager] asked the group to give quick reactions, most of the participants didn't speak or gesture as a response but used the UI feature showing a thumbs-up or smiley}{18}. The participants explained that they missed this feature on Engage and StarLeaf. On Engage, they found the meetings \say{less `affective' as I can't show any emotion with the icons}{13} as on AltspaceVR. 

\textbf{Lessons Learned.} Regardless of the primary purpose of meetings being of productive or social nature, the affective states of the attendees naturally have a critical impact on the collaboration. Whether for working or socializing with colleagues, it is essential to enable users to express their emotions and recognize those of others. This is especially pronounced with previously unknown people \citep{krieger2021.SocialEventinVR}. \cite{moustafa2018.longitudinalSocialVRGroups} found similar emotional states of users in social VR as in face-to-face settings that need to be considered. Still, our participants complained about the lack of facial expressions and other nonverbal communication, in line with the analysis by \cite{tanenbaum2020.ExpressiveNonverbalCommunicationinSVR}. We conjecture that the missing faces are a significant reason for the lower co-presence ratings of VR compared to video calls in our exploratory factor analysis. Considering the study by \cite{abdullah2021.videoconferenceVR}, this could also explain the participants' decreased activity to maintain social connectedness and less person-directed gaze in VR conversations compared to videoconferencing.
With rapid progress in research on the recognition of facial expressions \citep{lou_FacialReconstruction_2020, hamedi_facialExpressionRecog_2018, cha_fEMG_2022} and eye movement \citep{schwartz_eyes_2020}, as well as their mapping to the avatar, more expressive face and gaze representation will be available in the near future. We expect this to be a crucial facilitator for authentic social exchange in VR. Until then, a rudimentary way of sharing one's emotions or reactions is needed for successful interaction in VR, such as emojis. 
%If these reactions ``accurately reflect diverse user needs'' they can foster the exchange according to \cite{Cho2021.MoreThanThumbsUp}.

\subsection{Meeting Productivity}
As information exchange was a major aim of the meetings, productivity was a central theme in the data. When participants compared the VR meetings with in-person meetings or video calls, they often referred to them as \say{less productive}{05} and \say{inefficient, cumbersome}{10}. %, and \say{less effective}{01}. 
The reasons range from the inability to take notes and technical issues to the restrictions in seeing everyone simultaneously. Participants concluded that \say{the static meeting format}{18} of \say{regular reports and presentations do not benefit from VR}{05}. However, there were also some positive comments on the productivity of meetings in VR, as people listened \say{with less distractions and multitasking}{17} after a few AltspaceVR meetings. %While some participants disliked not being able to access secondary applications in VR, others enjoyed the benefit of the reduced distraction by these tools. 

\textbf{Attention and Distraction.} Being in VR reduced distractions from other applications like mail, web browsers, or instant messaging for our participants. At the same time, it introduced new distractions to the meeting as people explored the features: %\say{people are still playing around with the device and are detecting new features, which sometimes is distracting from the main task}{12}, 
\say{it can be distracting with all the non-meeting stuff you can do in VR}{14}. Overall there is no consensus on which medium is more beneficial regarding (in)attention, as some people expressed the need for a little diversion to focus on the meeting. 
Another aspect is the noticeable (in)attention of the meeting participants. Especially for the speaker, it is valuable feedback whether the audience actively listens. Some participants found that \say{in VR, one has a better awareness of the group, such as the closeness or distance, attention or distraction, etc., which is missing in Zoom.}{25}. In contrast, others thought \say{it was hard to tell [in VR] how distracted people were during the presentations. [...] In the physical meetings, it is more obvious who's listening and who's busy with something else}{18}. P11 had the impression that on Zoom \says{the participants were more focused and could follow the conversation better.}

\textbf{Presentations.} Giving and listening to presentations were reoccurring aspects. It felt for some participants more like an in-person setting. For the presenters, it was a more immersive and realistic experience than in video-based presentations: \say{I could walk in front of the screen and use my body. Felt like really presenting!}{18}. Similarly, P07 appreciated the \says{more true-to-life presentation environment with the projector screen and with the ability to see the audience spread out in a semicircle}. 
Nevertheless, for the audience, problems known from in-person presentations, such as view-blocking or display resolution, reduced satisfaction. Attendees reported a better view of the slides in videoconferences.

\textbf{Secondary tasks and tools. } %The aspects of this theme made up a big part of the data gathered in this study. As already mentioned, 
The access to notes had a substantial impact on the perceived productivity of the meeting, and comments like \say{I missed note taking}{28} for VR were very common. Similarly, access to tools like the calendar, search engines, and instant messengers for discreet parallel exchange within the team was missed. The lack of simultaneous messaging was mentioned frequently as it led to \say{less interactive}{17} meetings. In videoconferences, \say{most interactions occurred `outside the meeting' }{28} simultaneously via instant messaging. Several attendees mentioned that they could not do other productive tasks while being \say{trapped in VR}{28}, such as answering emails during personally irrelevant parts of the meeting. Also, the inability to access physical objects as beverages during VR sessions was criticized: \say{I miss my tea}{13} and \say{my coffee got cold}{18}.

\textbf{Lessons Learned.} For our work-focused meetings, the participants greatly appreciated the possibility of taking notes and accessing additional information. These aspects are not yet adequately provided by the VR tools tested in the scope of this study or require an unreasonable amount of individual effort. The accustomed workflow of a user should be supported in virtual meeting environments without allowing too many distracting elements into the VE. Recent developments promise a smoother integration of physical keyboards and applications from the desktop environment \citep{otte_keyboards_2019, oculus_Workrooms_2021}, which could help to bridge VEs and the usual workspace of a user. 

Although \cite{sarkar2021.VideoMeetingChats} found parallel chatting in video calls distracting and overwhelming for some users, they also identified valuable benefits from it that resonate with our participants' responses. More research needs to be done on integrating discreet side communication in VR. 
On the other hand, compared to video conferencing tools, participants valued the high-fidelity and engaging way of presenting in VR. This close resemblance matches the findings by \cite{ginkel2019.VRPresentationTraining} who measured similar skill improvement in an oral presentation training in VR as in a physical environment. But as similarly shown by \cite{yoshimura_study_2021}, the satisfaction with VR presentations also depended for our participants on the (dis)comfort with the equipment. Short and interactive talks can benefit more from VR than long or many presentations.

%\subsection{(In-)Convenience}
\subsection{User Needs}%Usability \& Users} % user needs? 
\label{sec:ResultsUserNeeds}
The primary objective of the attendees in our meetings was to get together once a week and exchange information. Their individual needs for functionality, comfort, and appropriate self-representation must be met.

\textbf{Technical Literacy of Users. } 
The questionnaire answers revealed some aspects rooted in the group's diversity. Although most of the participants had considerable knowledge of technology in general, some had serious technical challenges that sometimes even prevented them from participating at all, \say{because I wasn't able to set up the app}{19}, or required considerable effort: \say{I needed nearly an hour to join the meeting (and then just via [desktop PC]). And on top of that, I later needed help to get the VR setup running.}{25}. Even for experienced users, it took time to get familiar with the system as they encountered initial issues with the setup, updates, or controls. This clearly shows that even when using standard devices and commercially available applications, VR meetings still pose challenges to people who do not use the technology regularly. 
P09 was concerned about guests because \says{it's more of a hassle as they have to get accustomed to the software as well.} In particular, P23 worried about students presenting their thesis progress as it \says{could add to the pressure that they are already under and make them even more uncomfortable and stressed}.

\textbf{HMD discomfort.}
One frequently mentioned theme was the discomfort related to the VR headset. The participants reported motion sickness, headaches, that the \say{headset felt heavy and uncomfortable after some time in the meeting}{15}, and that it was \say{a bit stressful for my eyes}{12}. Participants used different HMD models, and some were unsatisfied with the visual quality. For example, P12 said about AltspaceVR: \says{the slides were difficult to read, the letters flickering; this was uncomfortable}. Furthermore, P17 mentioned that the Oculus Quest \says{got quite laggy with so many people. I had to reset the graphic settings. It gives me a headache and I'm more exhausted afterward.} 
But using the desktop client was also not a good alternative for some participants as they reported that \say{using 2D VR is a bit inconvenient for the control. It's somehow like [a first-person shooter], but I am not very used to that.}{13}

\textbf{Privacy. }
Some people described that they \say{felt observed}{12} on video calls because everybody could \say{see me and my living room}{18}. Regarding personal privacy and discretion, the VR sessions received positive feedback, as they allowed users to be invisible behind the avatar. Therefore, they did not need to worry about a presentable appearance or \say{need to show [their] facial expressions}{13}. Some participants also felt more confident as it \say{kinda helps with anxiety}{14}. 
However, users also reported that the spatial audio settings in AltspaceVR were insufficient for private conversations. P09 felt like \says{intruding} in private conversations: \says{After the meeting, it was weird as people talked in smaller groups, but one could hear everything}.

\textbf{Lessons Learned.} 
While some appreciated the concealment of personal appearance, emotional reactions, and the home office environment, private conversations could be heard from far away. After the effects of Zoom fatigue \citep{shockley2021.ZoomFatigue} during the pandemic, our participants felt relieved from the stress of being observed.
Although the quantitative results did not show any differences between the platforms in terms of factor \textit{F3 Privacy}, privacy was of concern in the qualitative data. The reason for this discrepancy is the different interpretations of the term. While standardized questionnaires and legal frameworks often consider privacy as the protection of information from the service provider, our participants were more concerned about the other attendees invading their privacy or being visually exposed to others. Quantitative methods should consider aspects of the personal privacy of meeting attendees beyond data protection regulations.

The fact that attendees from an HCI research group struggled with technical barriers clearly shows that meeting platforms must cater to users' diverse needs and abilities. Usability issues in many established social VR platforms, including AltspaceVR, were also identified by \cite{liu2021.UsabilitySocialVR}.  Using current off-the-shelf VR hardware and software resulted in a stressful experience with technical issues and HMD discomfort for some participants. Until the onboarding experience is smooth and quick for everyone, organizers must assume responsibility for providing support before the first meeting, as recommended similarly by \cite{lahlou2021.BeyondBeingThere}. It seems that VR meetings are currently not as inclusive and accessible as video-based platforms due to a higher technical entry barrier, even if expensive equipment is provided.

\subsection{Platform Choice for Future Meetings}
% alt titles:  Final Group Decision / Outcome of the Experiment /
In the final survey after the last meeting, we asked participants to draw a conclusion comparing the different platforms and to choose a long-term preference.
% pro VR

Many participants enjoyed exploring the possibility of meeting the team in VR: \say{it sure was fun and showed the potential for the future}{09}.
The users appreciated the possibilities of \say{free spatial arrangement}{12} with benefits especially for \say{engaging conversations between multiple actors}{05}  as it allows \say{spontaneously splitting into groups and chatting in clusters, without losing the overview of who else is present in the global room}{21}. 
Nevertheless, only two survey participants selected a VR platform as their preferred meeting medium.
P14 appreciated that it \says{is more fun and kinda helps with anxiety}. P18 supported using VR for future meetings because \says{on Zoom, it always feels like people can't wait to get out again. Most participants try to be as quiet as possible not to delay the meeting unnecessarily. In VR, there was much more socializing, arriving early and staying longer, personal and informal exchange, having fun, or enjoying meeting with others.}

% pro Zoom
However, the vast majority argued for returning to desktop-based video conferencing tools. Of the 15 participants who completed the survey after the last session, 13 preferred to go back to Zoom or StarLeaf. For most attendees, the \say{regular reports and presentations do not benefit from VR}{05} nor from the spatial component of Gather Town. Although fun, VEs were perceived as \say{not very helpful for serious work}{11}. P05 criticized the attempt \says{to replicate the reality without adding specific value that comes from the medium.}

On the contrary, for VR, the preparations and technical issues were disproportionately time-consuming, caused motion sickness for some attendees, secondary tasks were unavailable, and it was repeatedly described as inconvenient. For P19, the experience \says{was like speaking on an old phone with multiple participants at the same time, wearing a rock tied to my head, and watching some meaningless cartoon simultaneously}. On the other hand, in video calls, it is possible to see the faces of \say{people, and it works best. No preparations. No nausea or headache}{02}. It was described as \say{most convenient}{09} and \say{more productive}{13} with easy \say{access for external visitors}{11}. As many others, P12 \says{learned that seeing the faces of others is so important.}

\section{Limitations and Future Work}

We conducted this study in the wild during the regular team meetings of the research lab. Therefore, it has some limitations. Following an autoethnographic approach, the authors are part of the team and participated in the study. However, we ensured that the anonymization remained intact during coding and analysis. The study sample represents a very heterogeneous group as some team members have high technical literacy, and administrative staff was involved. This is most likely not representative of the general population, but for the domains that will be early adopters, this should be close to a representative sample. We also point out that most meeting attendees had no previous experience holding meetings in VR. Hence, the findings apply to teams of novice users and should only be extended to experts with caution. In addition, there was a trade-off between the length of the questionnaire and the aspects to cover, which influenced the direction of the results. 

To keep the atmosphere and behavior of the participants as natural as possible, we did not record the sessions. Consequently, no systematic observations of the attendees' behavior were possible, and we can only report on aspects described in the questionnaires. The selection, order, and frequency of the used platforms were team decisions independent of the study objectives, and therefore, the platforms were not explored evenly. Additionally, there might be possible biasing effects for the VR platform used second (Engage), as most users now had previous experiences with social VR. Furthermore, the content, length, and need for follow-up conversations varied every week due to circumstances, making it challenging to compare the meetings directly. During the meetings, various system configurations were used, which created different experiences for the individual users. We had a variety of HMDs (HTC Vive (Pro), Valve Index, and Oculus Quest), operating systems (Mac, Windows, Linux), and internet browsers (relevant for Gather Town). While this limits comparability between users, this case study was done in an authentic context with meetings that would have taken place in this manner regardless. All this ensures an authentic setting, leading to high external validity. 

Based on the outcome of this study, further research is needed with a clear separation of researchers and meeting attendees. At the cost of the insights from an autoethnographic approach, the generalizability can be higher with controlled, balanced, and diverse samples across domains and prior experience. Moreover, as a result of the independence of the meeting participants from the researchers, no negative consequences of honest answers are to be expected, which allows to lift the anonymity towards the data analysts and thus extends methodological possibilities. 
With a team that accepts the given parameters of the experiment, such as the choice of platforms and hardware or video recording of the meetings, future work could rely on more systematic and controlled circumstances enabling behavioral and linguistic analyses with high comparability. Still, a careful balance between external provisions and internal authenticity must be achieved.

\section{Conclusion}

% Summary
This in-the-wild case study on using social VR platforms and video conferencing software for weekly team meetings over the course of four months provides authentic insights into the experiences of 32 participants. Through questionnaires, we collected responses to open questions and Likert scale items after every meeting on various topics related to social interaction, meeting productivity, and individual experiences. In a thematic analysis of the qualitative data, we found five prominent themes that cover spatial aspects, meeting atmosphere, expression of emotions, meeting productivity, and user needs. The results show that regarding realistic group dynamics in gatherings before and after the official meeting, or for the impression of being together in the same place, VR meetings resemble in-person meetings more than videoconferences. Nevertheless, at the same time, videoconferences provided a closer-to-reality experience regarding seeing others' real faces and emotions, discreet side communication, and support of secondary tasks than VR meetings.

Furthermore, we performed a factor analysis on the quantitative data, revealing the advantages of videoconferences over VR platforms regarding ratings on involvement and co-presence. This finding is in line with the assessments in the final survey, in which participants were asked to draw a conclusion comparing the different platforms. Here, only 2 out of 15 survey participants selected a VR platform as their preferred meeting medium for weekly meetings in the future. Not being able to see the real faces of their colleagues was one of the most decisive factors for many participants. Therefore, at the end of the case study, the team decided to return to videoconferences for the long term. 
Some aspects found in this study have been discussed in related literature before. Still, this authentic in-the-wild setting of intrinsically motivated exploration of how suitable current commercial platforms are under natural working conditions enabled us to bring practical issues into context and highlight critical research gaps.

% Discussion: is social VR good enough, yet?
Our results indicate that the currently available off-the-shelf social VR platforms tested in this study do not yet sufficiently meet the needs and preferences of users to attend weekly team meetings like ours. Although they provide advantages in certain aspects, limitations such as time-consuming preparations, technical issues, HMD discomfort, motion sickness, and unavailable secondary tasks were among the main reasons for returning to video conferencing. Moreover, the findings indicate that the meeting format was not ideal for fully using the benefits of social VR. Above all, our results reveal how important it is for meeting attendees to see their colleagues' facial expressions and emotions and the current limitations of VR platforms in this regard. We suspect this to be the strongest contributor to the higher co-presence ratings of video calls compared to VR that we unexpectedly measured.

Some of the problems we encountered on our journey might be resolved within this decade thanks to technical advancements. As outlined in the Lessons Learned subsections above, new technologies using electromyogram and neuromuscular signals, as well as eye gaze and facial reconstruction, will provide avatars with rich emotional expressivity. Also, body language will be more realistic with advanced sensors and inverse kinematics. Peripherals such as physical keyboards and other components or applications of existing workflows will be seamlessly integrated into the virtual meeting environment. This considerably improves text input capabilities and the support of secondary tasks. In the meantime, usability issues on the software side will be resolved, and the devices will become lighter and more convenient.
Prospectively, if this case study had been conducted in a few years, the outcome concerning these challenges might be completely different. 

However, other challenges we encountered in our study might not be possible to overcome with technical progress and require careful consideration when (1) preparing technical aspects of the VE, and (2) planning and moderating the meeting. Considering our sample, these insights apply particularly to users who are new to having meetings in VR.
% Adapt VE
First, as part of the technical setup, an onboarding tutorial before the meeting must ensure that all attendants are familiar with the application and that their system is up-to-date. Users with previous experience should have had the opportunity to explore features in the VE before the meeting starts so it would not distract them. 
Further, the spatial audio should be configured for plenum situations in which speakers are audible by everyone anywhere but also cater for private (group) conversations with privacy-protecting sound bubbles and little distraction from distant background conversations. Acoustic parameters must always be transparent to users.
The application's overall visual style and professional appearance should match the seriousness of the occasion and the attendees' preferences. We recommend disabling permissions to insert distracting objects or special effects to the scene when not required for the meeting goal.

% Adapt Meeting Format
Second, to adapt the meeting format to the platform, a code of conduct should be agreed on to ensure a comfortable environment for everyone. Among other conventions, it should address proxemics, muting, recognizability, and modifications of the shared scene.
In line with related work, we found it important to designate a person responsible for technical support and facilitation, such as amplifying or muting attendees.
Generally, the format and goals of a meeting should be adapted to the advantages of social encounters in VR. Ideally, the reason for gathering is interpersonal interaction in dynamic group constellations with rich opportunities for socialization and creative exchange. Our participants appeared to appreciate situations before and after the meetings that afford the CoFIRe steps proposed by \cite{erickson2011.VRconferenceInteractions}: \textbf{co}alescence into small groups, \textbf{f}ocused \textbf{i}nteractions without distractions, and eventually \textbf{re}mixing with others. We recommend making use of spatial advantages and meaningful movements, such as forming queues, circles, and groups, flying when appropriate, moderating and taking turns with gaze and gestures, as well as organizing meeting content systematically within the space as demonstrated by \cite{luo_DocPlacement_2022} for augmented reality. 

We learned most of these lessons only as the study progressed. Initially, videoconferences were a substitute for in-person meetings due to the Covid-19 pandemic. The VR platforms were experimental substitutes for the new status quo of video conferencing. The reason for using any of these systems, however, was not to replicate in-person meetings but to achieve the purpose of the meetings.
How the group aimed to achieve these goals was organically shaped within the conditions of the respective platforms. Nevertheless, the meeting format was not sufficiently adapted to the benefits and restrictions of social VR platforms.
Technology does not break habits. And they appear to have been strong from the influence of a traditional meeting concept manifested over many years. The adjustment on-the-fly was not sufficient. Instead, a more conscious and targeted adaptation would have been required. 
Similarly, \cite{mcveigh-schultz_beyond_2021} argue for using XR to deliberately deviate from direct replication of face-to-face meetings, such as with augmentation of social behavior \citep{roth_beyondReplication_2018, roth_SocialAugmentations_2019} or ``social superpowers'' \citep{mcveigh-schultz_case_2021}.

A potential direction for the future could be a seamless integration of the different platform types to allow dynamic rearrangement according to the meeting situation, which often changes depending on the ongoing activities. This would allow using the most powerful medium for each case. For example, interactivity, small group interaction, poster sessions, or social events could take place in immersive XR environments, and detailed presentations or one-to-many announcements through video conferencing channels -- with immediate transition in between. Similar to the hybrid concept of Gather Town, this proposal attempts to combine the best of two solutions, although not limited to one device or a 2D world. 
%A first, yet two-dimensional, idea of such a setup can be found in tools like Gather Town or Wonder \footnote{\url{https://www.wonder.me/}}. 

% Wrap-Up
Overall, our case study provides authentic insights into conducting team meetings on off-the-shelf virtual reality platforms that can inform the appropriate choices and configurations of the platform, adaptations of the meeting format, and future requirements of social VR platforms.

%%%%%%%%%%%%%%%%%%%%%%%%%%%%%%%%%%%%%%%%%%%%%%%%%%%%%%%%%%%%%%%%%
%%%%%%%%%%%%% END OF BODY %%%%%%%%%%%%%%%%%%%%%%%%%&&%%%%%%%%%%%%
%%%%%%%%%%%%%%%%%%%%%%%%%%%%%%%%%%%%%%%%%%%%%%%%%%%%%%%%%%%%%%%%%

\section*{Conflict of Interest Statement}
%All financial, commercial or other relationships that might be perceived by the academic community as representing a potential conflict of interest must be disclosed. If no such relationship exists, authors will be asked to confirm the following statement: 

The authors declare that the research was conducted in the absence of any commercial or financial relationships that could be construed as a potential conflict of interest.

\section*{Author Contributions}
All authors developed the study design. MB prepared the test environments together with AR and coordinated the project. MB, AR, SP, YL, DA, RM, and TD carried out the data acquisition after the meetings. MB, AR, SP, YL, and TD were mainly responsible for data analysis. MB, AR, SP, and TD were responsible for drafting the manuscript. MB, AR, SP, and TD revised and finalized the manuscript.

%\section*{Funding}
%Details of all funding sources should be provided, including grant numbers if applicable. Please ensure to add all necessary funding information, as after publication this is no longer possible.

\section*{Acknowledgments}
The Klaus Tschira Foundation partially funded this research. We want to thank all team members for their commitment, feedback, and patience in this study. Also, we want to thank the Engage team, who provided us with a free trial for larger groups to run our study.
A preprint of this paper is available on arXiv.org \citep{ThisPaperPreprint}. 

\section*{Supplemental Data}
\label{supplements}
The supplementary material to this article includes a video figure, the questionnaires used in the study, an overview of the questionnaire items, and additional screenshots of the meetings in VR.

\section*{Data Availability Statement}
The quantitative dataset presented in this article can be found at \url{https://doi.org/10.17605/OSF.IO/YG7HK}.
The qualitative dataset presented in this article is not publicly available because the participants in the case study can easily be identified by their affiliation. Therefore, we assure them of the complete confidentiality of the collected data. Only the authors, excluding the group manager, could access the data to protect participants' privacy and encourage them to respond authentically without fear of repercussions.

% Please see the availability of data guidelines for more information, at https://www.frontiersin.org/about/author-guidelines#AvailabilityofData

\bibliographystyle{frontiersinSCNS_ENG_HUMS} % for Science, Engineering and Humanities and Social Sciences articles, for Humanities and Social Sciences articles please include page numbers in the in-text citations
\bibliography{references}

\begin{thebibliography}{85}
\providecommand{\natexlab}[1]{#1}
\expandafter\ifx\csname urlstyle\endcsname\relax
  \providecommand{\doi}[1]{doi:\discretionary{}{}{}#1}\else
  \providecommand{\doi}{doi:\discretionary{}{}{}\begingroup
  \urlstyle{rm}\Url}\fi
\providecommand{\selectlanguage}[1]{\relax}
\providecommand{\bibAnnoteFile}[1]{%
  \IfFileExists{#1}{\begin{quotation}\noindent\textsc{Key:} #1\\
  \textsc{Annotation:}\ \input{#1}\end{quotation}}{}}
\providecommand{\bibAnnote}[2]{%
  \begin{quotation}\noindent\textsc{Key:} #1\\
  \textsc{Annotation:}\ #2\end{quotation}}

\bibitem[{Abdullah et~al.(2021)Abdullah, Kolkmeier, Lo, and
  Neff}]{abdullah2021.videoconferenceVR}
Abdullah, A., Kolkmeier, J., Lo, V., and Neff, M. (2021).
\newblock Videoconference and {Embodied} {VR}: {Communication} {Patterns}
  {Across} {Task} and {Medium}.
\newblock \emph{Proceedings of the ACM on Human-Computer Interaction} 5,
  453:1--453:29.
\newblock \doi{10.1145/3479597}
\bibAnnoteFile{abdullah2021.videoconferenceVR}

\bibitem[{Ahn et~al.(2021)Ahn, Levy, Eden, Won, {MacIntyre}, and
  Johnsen}]{ahn_ieeevr2020_2021}
Ahn, S. J.~G., Levy, L., Eden, A., Won, A.~S., {MacIntyre}, B., and Johnsen, K.
  (2021).
\newblock {IEEEVR}2020: Exploring the first steps toward standalone virtual
  conferences 2.
\newblock \doi{10.3389/frvir.2021.648575}
\bibAnnoteFile{ahn_ieeevr2020_2021}

\bibitem[{Bailenson(2021)}]{bailenson_nonverbal_2021}
Bailenson, J.~N. (2021).
\newblock Nonverbal {Overload}: {A} {Theoretical} {Argument} for the {Causes}
  of {Zoom} {Fatigue}.
\newblock \emph{Technology, Mind, and Behavior} 2.
\newblock \doi{10.1037/tmb0000030}
\bibAnnoteFile{bailenson_nonverbal_2021}

\bibitem[{Bailenson et~al.(2004)Bailenson, Beall, Loomis, Blascovich, and
  Turk}]{bailenson_transformed_2004}
Bailenson, J.~N., Beall, A.~C., Loomis, J., Blascovich, J., and Turk, M.
  (2004).
\newblock Transformed social interaction: Decoupling representation from
  behavior and form in collaborative virtual environments 13, 428--441.
\newblock \doi{10.1162/1054746041944803}
\bibAnnoteFile{bailenson_transformed_2004}

\bibitem[{Bailenson et~al.(2001)Bailenson, Blascovich, Beall, and
  Loomis}]{bailenson_equilibriumPersonalSpace_2001}
Bailenson, J.~N., Blascovich, J., Beall, A.~C., and Loomis, J.~M. (2001).
\newblock Equilibrium {Theory} {Revisited}: {Mutual} {Gaze} and {Personal}
  {Space} in {Virtual} {Environments}.
\newblock \emph{Presence: Teleoperators and Virtual Environments} 10, 583--598.
\newblock \doi{10.1162/105474601753272844}
\bibAnnoteFile{bailenson_equilibriumPersonalSpace_2001}

\bibitem[{Bailenson et~al.(2005)Bailenson, Swinth, Hoyt, Persky, Dimov, and
  Blascovich}]{bailenson_independent_2005}
Bailenson, J.~N., Swinth, K., Hoyt, C., Persky, S., Dimov, A., and Blascovich,
  J. (2005).
\newblock The {Independent} and {Interactive} {Effects} of {Embodied}-{Agent}
  {Appearance} and {Behavior} on {Self}-{Report}, {Cognitive}, and {Behavioral}
  {Markers} of {Copresence} in {Immersive} {Virtual} {Environments}.
\newblock \emph{Presence: Teleoperators and Virtual Environments} 14, 379--393.
\newblock \doi{10.1162/105474605774785235}
\bibAnnoteFile{bailenson_independent_2005}

\bibitem[{Barreda-Ángeles and
  Hartmann(2022)}]{Barreda-Angeles2022.PsychoEffectsSocialVRPresence}
Barreda-Ángeles, M. and Hartmann, T. (2022).
\newblock Psychological benefits of using social virtual reality platforms
  during the covid-19 pandemic: The role of social and spatial presence.
\newblock \emph{Computers in Human Behavior} 127, 107047.
\newblock \doi{https://doi.org/10.1016/j.chb.2021.107047}
\bibAnnoteFile{Barreda-Angeles2022.PsychoEffectsSocialVRPresence}

\bibitem[{Bente et~al.(2008)Bente, Rüggenberg, Krämer, and
  Eschenburg}]{bente_noImpactAvatarRealism_2008}
Bente, G., Rüggenberg, S., Krämer, N.~C., and Eschenburg, F. (2008).
\newblock Avatar-{Mediated} {Networking}: {Increasing} {Social} {Presence} and
  {Interpersonal} {Trust} in {Net}-{Based} {Collaborations}.
\newblock \emph{Human Communication Research} 34, 287--318.
\newblock \doi{10.1111/j.1468-2958.2008.00322.x}
\bibAnnoteFile{bente_noImpactAvatarRealism_2008}

\bibitem[{Blackwell et~al.(2019{\natexlab{a}})Blackwell, Ellison,
  Elliott-Deflo, and Schwartz}]{blackwell_harassment-governance_2019}
Blackwell, L., Ellison, N., Elliott-Deflo, N., and Schwartz, R.
  (2019{\natexlab{a}}).
\newblock Harassment in {Social} {Virtual} {Reality}: {Challenges} for
  {Platform} {Governance}.
\newblock \emph{Proceedings of the ACM on Human-Computer Interaction} 3,
  100:1--100:25.
\newblock \doi{10.1145/3359202}
\bibAnnoteFile{blackwell_harassment-governance_2019}

\bibitem[{Blackwell et~al.(2019{\natexlab{b}})Blackwell, Ellison,
  Elliott-Deflo, and Schwartz}]{blackwell_harassment_2019}
Blackwell, L., Ellison, N., Elliott-Deflo, N., and Schwartz, R.
  (2019{\natexlab{b}}).
\newblock Harassment in {Social} {VR}: {Implications} for {Design}.
\newblock In \emph{2019 {IEEE} {Conference} on {Virtual} {Reality} and {3D}
  {User} {Interfaces} ({VR})}. 854--855.
\newblock \doi{10.1109/VR.2019.8798165}.
\newblock ISSN: 2642-5254
\bibAnnoteFile{blackwell_harassment_2019}

\bibitem[{Bleakley et~al.(2022)Bleakley, Rough, Edwards, Doyle, Dumbleton,
  Clark et~al.}]{bleakley2022.SocialTalk}
Bleakley, A., Rough, D., Edwards, J., Doyle, P., Dumbleton, O., Clark, L.,
  et~al. (2022).
\newblock Bridging social distance during social distancing: exploring social
  talk and remote collegiality in video conferencing.
\newblock \emph{Human–Computer Interaction} 37, 404--432.
\newblock \doi{10.1080/07370024.2021.1994859}.
\newblock Publisher: Taylor \& Francis \_eprint:
  https://doi.org/10.1080/07370024.2021.1994859
\bibAnnoteFile{bleakley2022.SocialTalk}

\bibitem[{Bonfert et~al.(2022)Bonfert, Reinschluessel, Putze, Lai,
  Alexandrovsky, Malaka et~al.}]{ThisPaperPreprint}
Bonfert, M., Reinschluessel, A.~V., Putze, S., Lai, Y., Alexandrovsky, D.,
  Malaka, R., et~al. (2022).
\newblock "{Seeing} the {Faces} {Is} {So} {Important}" -- {Experiences} {From}
  {Online} {Team} {Meetings} on {Commercial} {Virtual} {Reality} {Platforms}
  \doi{10.48550/arXiv.2210.06190}.
\newblock Preprint.
\bibAnnoteFile{ThisPaperPreprint}

\bibitem[{Brucks and Levav(2022)}]{Brucks2022ve}
Brucks, M.~S. and Levav, J. (2022).
\newblock Virtual communication curbs creative idea generation.
\newblock \emph{Nature}
\bibAnnoteFile{Brucks2022ve}

\bibitem[{Burgoon et~al.(1999)Burgoon, Bonito, Bengtsson, Ramirez, Dunbar, and
  Miczo}]{burgoon_testing_1999}
Burgoon, J.~K., Bonito, J.~A., Bengtsson, B., Ramirez, A., Dunbar, N.~E., and
  Miczo, N. (1999).
\newblock Testing the {Interactivity} {Model}: {Communication} {Processes},
  {Partner} {Assessments}, and the {Quality} of {Collaborative} {Work}.
\newblock \emph{Journal of Management Information Systems} 16, 33--56.
\newblock \doi{10.1080/07421222.1999.11518255}.
\newblock Publisher: Routledge \_eprint:
  https://doi.org/10.1080/07421222.1999.11518255
\bibAnnoteFile{burgoon_testing_1999}

\bibitem[{Byun et~al.(2011)Byun, Awasthi, Chou, Kapoor, Lee, and
  Czerwinski}]{byun2011.HonestSignals}
Byun, B., Awasthi, A., Chou, P.~A., Kapoor, A., Lee, B., and Czerwinski, M.
  (2011).
\newblock Honest signals in video conferencing.
\newblock In \emph{2011 {IEEE} {International} {Conference} on {Multimedia} and
  {Expo}}. 1--6.
\newblock \doi{10.1109/ICME.2011.6011855}.
\newblock ISSN: 1945-788X
\bibAnnoteFile{byun2011.HonestSignals}

\bibitem[{Cha and Im(2022)}]{cha_fEMG_2022}
Cha, H.-S. and Im, C.-H. (2022).
\newblock Performance enhancement of facial electromyogram-based
  facial-expression recognition for social virtual reality applications using
  linear discriminant analysis adaptation.
\newblock \emph{Virtual Reality} 26, 385--398.
\newblock \doi{10.1007/s10055-021-00575-6}
\bibAnnoteFile{cha_fEMG_2022}

\bibitem[{Daft and Lengel(1986)}]{daft_organizational_1986}
Daft, R.~L. and Lengel, R.~H. (1986).
\newblock Organizational {Information} {Requirements}, {Media} {Richness} and
  {Structural} {Design}.
\newblock \emph{Management Science} 32, 554--571.
\newblock \doi{10.1287/mnsc.32.5.554}.
\newblock Publisher: INFORMS
\bibAnnoteFile{daft_organizational_1986}

\bibitem[{Dzardanova et~al.(2021)Dzardanova, Kasapakis, Gavalas, and
  Sylaiou}]{dzardanova_virtual_2021}
Dzardanova, E., Kasapakis, V., Gavalas, D., and Sylaiou, S. (2021).
\newblock Virtual reality as a communication medium: a comparative study of
  forced compliance in virtual reality versus physical world
  \doi{10.1007/s10055-021-00564-9}
\bibAnnoteFile{dzardanova_virtual_2021}

\bibitem[{Erickson et~al.(2011)Erickson, Shami, Kellogg, and
  Levine}]{erickson2011.VRconferenceInteractions}
Erickson, T., Shami, N.~S., Kellogg, W.~A., and Levine, D.~W. (2011).
\newblock Synchronous interaction among hundreds: an evaluation of a conference
  in an avatar-based virtual environment.
\newblock In \emph{Proceedings of the {SIGCHI} {Conference} on {Human}
  {Factors} in {Computing} {Systems}} (New York, NY, USA: Association for
  Computing Machinery), {CHI} '11, 503--512.
\newblock \doi{10.1145/1978942.1979013}
\bibAnnoteFile{erickson2011.VRconferenceInteractions}

\bibitem[{Finn et~al.(1997)Finn, Sellen, and Wilbur}]{finn_video-mediated_1997}
Finn, K.~E., Sellen, A.~J., and Wilbur, S. (1997).
\newblock \emph{Video-mediated {Communication}} (L. Erlbaum Associates).
\newblock Google-Books-ID: T35HAAAAMAAJ
\bibAnnoteFile{finn_video-mediated_1997}

\bibitem[{Freeman et~al.(2022)Freeman, Zamanifard, Maloney, and
  Acena}]{Freeman2022.VRHarassment}
Freeman, G., Zamanifard, S., Maloney, D., and Acena, D. (2022).
\newblock Disturbing the peace: Experiencing and mitigating emerging harassment
  in social virtual reality.
\newblock \emph{Proc. ACM Hum.-Comput. Interact.} 6.
\newblock \doi{10.1145/3512932}
\bibAnnoteFile{Freeman2022.VRHarassment}

\bibitem[{Fägersten(2010)}]{fagersten2010.DiscourseAnalysis}
Fägersten, K. (2010).
\newblock Using discourse analysis to assess social co-presence in the video
  conference environment.
\newblock In \emph{Cases on {Online} {Discussion} and {Interaction}:
  {Experiences} and {Outcomes}} (Hershey, PA: IGI Global). 19
\bibAnnoteFile{fagersten2010.DiscourseAnalysis}

\bibitem[{Galegher et~al.(2013)Galegher, Kraut, and
  Egido}]{galegher_intellectual-teamwork_2013}
Galegher, J., Kraut, R.~E., and Egido, C. (eds.) (2013).
\newblock \emph{Intellectual {Teamwork}: {Social} and {Technological}
  {Foundations} of {Cooperative} {Work}} (New York: Psychology Press).
\newblock \doi{10.4324/9781315807645}
\bibAnnoteFile{galegher_intellectual-teamwork_2013}

\bibitem[{Ginkel et~al.(2019)Ginkel, Gulikers, Biemans, Noroozi, Roozen, Bos
  et~al.}]{ginkel2019.VRPresentationTraining}
Ginkel, S.~v., Gulikers, J., Biemans, H., Noroozi, O., Roozen, M., Bos, T.,
  et~al. (2019).
\newblock Fostering oral presentation competence through a virtual
  reality-based task for delivering feedback.
\newblock \emph{Computers \& Education} 134, 78--97.
\newblock \doi{https://doi.org/10.1016/j.compedu.2019.02.006}
\bibAnnoteFile{ginkel2019.VRPresentationTraining}

\bibitem[{Gonzalez-Franco et~al.(2017)Gonzalez-Franco, Maselli, Florencio,
  Smolyanskiy, and Zhang}]{gonzalez-franco_concurrentTalking_2017}
Gonzalez-Franco, M., Maselli, A., Florencio, D., Smolyanskiy, N., and Zhang, Z.
  (2017).
\newblock Concurrent talking in immersive virtual reality: on the dominance of
  visual speech cues.
\newblock \emph{Scientific Reports} 7, 3817.
\newblock \doi{10.1038/s41598-017-04201-x}.
\newblock Number: 1 Publisher: Nature Publishing Group
\bibAnnoteFile{gonzalez-franco_concurrentTalking_2017}

\bibitem[{Hamedi et~al.(2018)Hamedi, Salleh, Ting, Astaraki, and
  Noor}]{hamedi_facialExpressionRecog_2018}
Hamedi, M., Salleh, S.-H., Ting, C.-M., Astaraki, M., and Noor, A.~M. (2018).
\newblock Robust {Facial} {Expression} {Recognition} for {MuCI}: {A}
  {Comprehensive} {Neuromuscular} {Signal} {Analysis}.
\newblock \emph{IEEE Transactions on Affective Computing} 9, 102--115.
\newblock \doi{10.1109/TAFFC.2016.2569098}.
\newblock Conference Name: IEEE Transactions on Affective Computing
\bibAnnoteFile{hamedi_facialExpressionRecog_2018}

\bibitem[{Hecht et~al.(2019)Hecht, Welsch, Viehoff, and
  Longo}]{hecht_shapePersonalSpace_2019}
Hecht, H., Welsch, R., Viehoff, J., and Longo, M.~R. (2019).
\newblock The shape of personal space.
\newblock \emph{Acta Psychologica} 193, 113--122.
\newblock \doi{10.1016/j.actpsy.2018.12.009}
\bibAnnoteFile{hecht_shapePersonalSpace_2019}

\bibitem[{Hinds(1999)}]{hinds_cognitive_1999}
Hinds, P.~J. (1999).
\newblock The {Cognitive} and {Interpersonal} {Costs} of {Video}.
\newblock \emph{Media Psychology} 1, 283--311.
\newblock \doi{10.1207/s1532785xmep0104_1}.
\newblock Publisher: Routledge \_eprint:
  https://doi.org/10.1207/s1532785xmep0104\_1
\bibAnnoteFile{hinds_cognitive_1999}

\bibitem[{Kang and Watt(2013)}]{kang_impactAvatarRealism_2013}
Kang, S.-H. and Watt, J.~H. (2013).
\newblock The impact of avatar realism and anonymity on effective communication
  via mobile devices.
\newblock \emph{Computers in Human Behavior} 29, 1169--1181.
\newblock \doi{10.1016/j.chb.2012.10.010}
\bibAnnoteFile{kang_impactAvatarRealism_2013}

\bibitem[{Kern and Ellermeier(2020)}]{kern_VRaudio_2020}
Kern, A.~C. and Ellermeier, W. (2020).
\newblock Audio in {VR}: {Effects} of a {Soundscape} and {Movement}-{Triggered}
  {Step} {Sounds} on {Presence}.
\newblock \emph{Frontiers in Robotics and AI} 7
\bibAnnoteFile{kern_VRaudio_2020}

\bibitem[{Kirchner and Nordin~Forsberg(2021)}]{krieger2021.SocialEventinVR}
Kirchner, K. and Nordin~Forsberg, B. (2021).
\newblock A {Conference} {Goes} {Virtual}: {Lessons} from {Creating} a {Social}
  {Event} in the {Virtual} {Reality}.
\newblock In \emph{Innovations for {Community} {Services}}, eds. U.~R. Krieger,
  G.~Eichler, C.~Erfurth, and G.~Fahrnberger (Cham: Springer International
  Publishing), vol. 1404. 123--134.
\newblock \doi{10.1007/978-3-030-75004-6_9}.
\newblock 00000 Series Title: Communications in Computer and Information
  Science
\bibAnnoteFile{krieger2021.SocialEventinVR}

\bibitem[{Koseki et~al.(2020)Koseki, Iijima, Yamamoto, and
  Seiki}]{Koseki2020.CommProblems}
Koseki, N., Iijima, M., Yamamoto, Y., and Seiki, K. (2020).
\newblock Identifying communication problems in the use of video conferencing
  systems.
\newblock In \emph{2020 9th International Congress on Advanced Applied
  Informatics (IIAI-AAI)}. 651--658.
\newblock \doi{10.1109/IIAI-AAI50415.2020.00133}
\bibAnnoteFile{Koseki2020.CommProblems}

\bibitem[{Kramer et~al.(2006)Kramer, Oh, and
  Fussell}]{Kramer2006.LinguisticPresence}
Kramer, A. D.~I., Oh, L.~M., and Fussell, S.~R. (2006).
\newblock Using linguistic features to measure presence in computer-mediated
  communication.
\newblock In \emph{Proceedings of the SIGCHI Conference on Human Factors in
  Computing Systems} (New York, NY, USA: Association for Computing Machinery),
  CHI '06, 913–916.
\newblock \doi{10.1145/1124772.1124907}
\bibAnnoteFile{Kramer2006.LinguisticPresence}

\bibitem[{Kyrlitsias and Michael-Grigoriou(2022)}]{kyrlitsias_social_2022}
Kyrlitsias, C. and Michael-Grigoriou, D. (2022).
\newblock Social interaction with agents and avatars in immersive virtual
  environments: A survey 2.
\newblock \doi{10.3389/frvir.2021.786665}
\bibAnnoteFile{kyrlitsias_social_2022}

\bibitem[{Lahlou et~al.(2021)Lahlou, Pea, Heitmayer, G.~Russell,
  Schimmelpfennig, Yamin et~al.}]{lahlou2021.BeyondBeingThere}
Lahlou, S., Pea, R., Heitmayer, M., G.~Russell, M., Schimmelpfennig, R., Yamin,
  P., et~al. (2021).
\newblock Are we ‘{Beyond} being there’ yet?: {Towards} better interweaving
  epistemic and social aspects of virtual reality conferencing.
\newblock In \emph{Extended {Abstracts} of the 2021 {CHI} {Conference} on
  {Human} {Factors} in {Computing} {Systems}} (Yokohama Japan: ACM), 1--6.
\newblock \doi{10.1145/3411763.3451579}.
\newblock 00000
\bibAnnoteFile{lahlou2021.BeyondBeingThere}

\bibitem[{Liu and Steed(2021)}]{liu2021.UsabilitySocialVR}
Liu, Q. and Steed, A. (2021).
\newblock Social {Virtual} {Reality} {Platform} {Comparison} and {Evaluation}
  {Using} a {Guided} {Group} {Walkthrough} {Method}.
\newblock \emph{Frontiers in Virtual Reality} 2, 668181.
\newblock \doi{10.3389/frvir.2021.668181}.
\newblock 00000
\bibAnnoteFile{liu2021.UsabilitySocialVR}

\bibitem[{Lou et~al.(2020)Lou, Wang, Nduka, Hamedi, Mavridou, Wang
  et~al.}]{lou_FacialReconstruction_2020}
Lou, J., Wang, Y., Nduka, C., Hamedi, M., Mavridou, I., Wang, F.-Y., et~al.
  (2020).
\newblock Realistic {Facial} {Expression} {Reconstruction} for {VR} {HMD}
  {Users}.
\newblock \emph{IEEE Transactions on Multimedia} 22, 730--743.
\newblock \doi{10.1109/TMM.2019.2933338}.
\newblock Conference Name: IEEE Transactions on Multimedia
\bibAnnoteFile{lou_FacialReconstruction_2020}

\bibitem[{Luo et~al.(2022)Luo, Lehmann, Widengren, and
  Dachselt}]{luo_DocPlacement_2022}
Luo, W., Lehmann, A., Widengren, H., and Dachselt, R. (2022).
\newblock Where {Should} {We} {Put} {It}? {Layout} and {Placement} {Strategies}
  of {Documents} in {Augmented} {Reality} for {Collaborative} {Sensemaking}.
\newblock In \emph{{CHI} {Conference} on {Human} {Factors} in {Computing}
  {Systems}} (New York, NY, USA: Association for Computing Machinery), {CHI}
  '22, 1--16.
\newblock \doi{10.1145/3491102.3501946}
\bibAnnoteFile{luo_DocPlacement_2022}

\bibitem[{Mack et~al.(2021)Mack, Das, Jain, Bragg, Tang, Begel
  et~al.}]{Mack2021.AutoethnographicInterns}
Mack, K., Das, M., Jain, D., Bragg, D., Tang, J., Begel, A., et~al. (2021).
\newblock Mixed {Abilities} and {Varied} {Experiences}: a group autoethnography
  of a virtual summer internship.
\newblock In \emph{The 23rd {International} {ACM} {SIGACCESS} {Conference} on
  {Computers} and {Accessibility}} (New York, NY, USA: Association for
  Computing Machinery), {ASSETS} '21, 1--13.
\newblock \doi{10.1145/3441852.3471199}
\bibAnnoteFile{Mack2021.AutoethnographicInterns}

\bibitem[{Maloney et~al.(2020)Maloney, Freeman, and
  Wohn}]{maloney_talking_2020}
Maloney, D., Freeman, G., and Wohn, D.~Y. (2020).
\newblock "talking without a voice": Understanding non-verbal communication in
  social virtual reality 4, 175:1--175:25.
\newblock \doi{10.1145/3415246}
\bibAnnoteFile{maloney_talking_2020}

\bibitem[{{McVeigh}-Schultz and
  Isbister(2021{\natexlab{a}})}]{mcveigh-schultz_case_2021}
{McVeigh}-Schultz, J. and Isbister, K. (2021{\natexlab{a}}).
\newblock The case for “weird social” in {VR}/{XR}: A vision of social
  superpowers beyond meatspace.
\newblock In \emph{Extended Abstracts of the 2021 {CHI} Conference on Human
  Factors in Computing Systems} (Association for Computing Machinery), 17.
  1--10
\bibAnnoteFile{mcveigh-schultz_case_2021}

\bibitem[{{McVeigh}-Schultz and
  Isbister(2021{\natexlab{b}})}]{mcveigh-schultz_beyond_2021}
{McVeigh}-Schultz, J. and Isbister, K. (2021{\natexlab{b}}).
\newblock A “beyond being there” for {VR} meetings: envisioning the future
  of remote work 0, 1--21.
\newblock \doi{10.1080/07370024.2021.1994860}.
\newblock 00001 Publisher: Taylor \& Francis \_eprint:
  https://doi.org/10.1080/07370024.2021.1994860
\bibAnnoteFile{mcveigh-schultz_beyond_2021}

\bibitem[{McVeigh-Schultz et~al.(2019)McVeigh-Schultz, Kolesnichenko, and
  Isbister}]{mcveigh-schultz_pro-social_2019}
McVeigh-Schultz, J., Kolesnichenko, A., and Isbister, K. (2019).
\newblock Shaping {Pro}-{Social} {Interaction} in {VR}: {An} {Emerging}
  {Design} {Framework}.
\newblock In \emph{Proceedings of the 2019 {CHI} {Conference} on {Human}
  {Factors} in {Computing} {Systems}} (New York, NY, USA: Association for
  Computing Machinery), {CHI} '19, 1--12.
\newblock \doi{10.1145/3290605.3300794}
\bibAnnoteFile{mcveigh-schultz_pro-social_2019}

\bibitem[{McVeigh-Schultz et~al.(2018)McVeigh-Schultz, Márquez~Segura,
  Merrill, and Isbister}]{mcveigh-schultz_whats_2018}
McVeigh-Schultz, J., Márquez~Segura, E., Merrill, N., and Isbister, K. (2018).
\newblock What's {It} {Mean} to "{Be} {Social}" in {VR}? {Mapping} the {Social}
  {VR} {Design} {Ecology}.
\newblock In \emph{Proceedings of the 2018 {ACM} {Conference} {Companion}
  {Publication} on {Designing} {Interactive} {Systems}} (New York, NY, USA:
  Association for Computing Machinery), {DIS} '18 {Companion}, 289--294.
\newblock \doi{10.1145/3197391.3205451}
\bibAnnoteFile{mcveigh-schultz_whats_2018}

\bibitem[{Microsoft(2022)}]{microsoft2022.WorkTrendIndex}
[Dataset] Microsoft (2022).
\newblock Work trend index 2022. great expectations: Making hybrid work work
\bibAnnoteFile{microsoft2022.WorkTrendIndex}

\bibitem[{Milgram and Gudehus(1978)}]{milgram_obedience_1978}
[Dataset] Milgram, S. and Gudehus, C. (1978).
\newblock Obedience to authority
\bibAnnoteFile{milgram_obedience_1978}

\bibitem[{Moustafa and Steed(2018)}]{moustafa2018.longitudinalSocialVRGroups}
Moustafa, F. and Steed, A. (2018).
\newblock A longitudinal study of small group interaction in social virtual
  reality.
\newblock In \emph{Proceedings of the 24th {ACM} {Symposium} on {Virtual}
  {Reality} {Software} and {Technology}} (New York, NY, USA: Association for
  Computing Machinery), {VRST} '18, 1--10.
\newblock \doi{10.1145/3281505.3281527}
\bibAnnoteFile{moustafa2018.longitudinalSocialVRGroups}

\bibitem[{Nadler(2020)}]{nadler_understanding_2020}
Nadler, R. (2020).
\newblock Understanding “{Zoom} fatigue”: {Theorizing} spatial dynamics as
  third skins in computer-mediated communication.
\newblock \emph{Computers and Composition} 58, 102613.
\newblock \doi{10.1016/j.compcom.2020.102613}
\bibAnnoteFile{nadler_understanding_2020}

\bibitem[{Nilles(1975)}]{nilles_telecommunications_1975}
Nilles, J. (1975).
\newblock Telecommunications and {Organizational} {Decentralization}.
\newblock \emph{IEEE Transactions on Communications} 23, 1142--1147.
\newblock \doi{10.1109/TCOM.1975.1092687}.
\newblock Conference Name: IEEE Transactions on Communications
\bibAnnoteFile{nilles_telecommunications_1975}

\bibitem[{Nind et~al.(2021)Nind, Coverdale, and
  Meckin}]{Nind2021.ResearchPracticesCovid}
Nind, M., Coverdale, A., and Meckin, R. (2021).
\newblock \emph{Changing {Social} {Research} {Practices} in the {Context} of
  {Covid}-19: {Rapid} {Evidence} {Review}}.
\newblock Working {Paper}, NCRM.
\newblock \doi{10.5258/NCRM/NCRM.00004458}
\bibAnnoteFile{Nind2021.ResearchPracticesCovid}

\bibitem[{Nowak(2001)}]{nowak_Social-Co-Presence_2001}
Nowak, K. (2001).
\newblock Defining and {Differentiating} {Copresence}, {Social} {Presence} and
  {Presence} as {Transportation}.
\newblock In \emph{4 th {Annual} {International} {Workshop}, {Philadelphia},
  {Rettie}, {R}.{M}., 2003, {A} {Comparison} of {Four} {New} {Communication}
  {Technologies}, {Proceedings} of {HCI} {International} {Conference} on
  {Human}-{Computer} {Interaction}, {Lawrence} {Erlbaum} {Associates}}.
  686--690
\bibAnnoteFile{nowak_Social-Co-Presence_2001}

\bibitem[{Nowak and Biocca(2003)}]{Nowak2003.SocialPresence}
Nowak, K.~L. and Biocca, F. (2003).
\newblock {The Effect of the Agency and Anthropomorphism on Users' Sense of
  Telepresence, Copresence, and Social Presence in Virtual Environments}.
\newblock \emph{Presence: Teleoperators and Virtual Environments} 12, 481--494.
\newblock \doi{10.1162/105474603322761289}
\bibAnnoteFile{Nowak2003.SocialPresence}

\bibitem[{Ochsman and Chapanis(1974)}]{ochsman_effects_1974}
Ochsman, R.~B. and Chapanis, A. (1974).
\newblock The {Effects} of 10 {Communication} {Modes} on the {Behavior} of
  {Teams} {During} {Co}-{Operative} {Problem}-{Solving}.
\newblock \emph{Int. J. Man Mach. Stud.} \doi{10.1016/S0020-7373(74)80019-2}
\bibAnnoteFile{ochsman_effects_1974}

\bibitem[{Oculus(2021)}]{oculus_Workrooms_2021}
[Dataset] Oculus (2021).
\newblock Introducing ‘{Horizon} {Workrooms}’: {Remote} {Collaboration}
  {Reimagined}
\bibAnnoteFile{oculus_Workrooms_2021}

\bibitem[{OECD(2021)}]{oecd_teleworking_2021}
OECD (2021).
\newblock \emph{Teleworking in the {COVID}-19 {Pandemic}: {Trends} and
  {Prospects}}.
\newblock Tech. rep., Organisation for Economic Co-operation and Development
\bibAnnoteFile{oecd_teleworking_2021}

\bibitem[{Oh et~al.(2018)Oh, Bailenson, and
  Welch}]{oh_systematicSocialPresence_2018}
Oh, C.~S., Bailenson, J.~N., and Welch, G.~F. (2018).
\newblock A {Systematic} {Review} of {Social} {Presence}: {Definition},
  {Antecedents}, and {Implications}.
\newblock \emph{Frontiers in Robotics and AI} 5
\bibAnnoteFile{oh_systematicSocialPresence_2018}

\bibitem[{Olson et~al.(1997)Olson, Olson, and Meader}]{olson_face-to-face_1997}
Olson, J.~S., Olson, G.~M., and Meader, D. (1997).
\newblock Face-to-face group work compared to remote group work with and
  without video.
\newblock In \emph{Video-mediated communication} (Mahwah, NJ, US: Lawrence
  Erlbaum Associates Publishers), Computers, cognition, and work. 157--172
\bibAnnoteFile{olson_face-to-face_1997}

\bibitem[{Otte et~al.(2019)Otte, Menzner, Gesslein, Gagel, Schneider, and
  Grubert}]{otte_keyboards_2019}
Otte, A., Menzner, T., Gesslein, T., Gagel, P., Schneider, D., and Grubert, J.
  (2019).
\newblock Towards {Utilizing} {Touch}-sensitive {Physical} {Keyboards} for
  {Text} {Entry} in {Virtual} {Reality}.
\newblock In \emph{2019 {IEEE} {Conference} on {Virtual} {Reality} and {3D}
  {User} {Interfaces} ({VR})}. 1729--1732.
\newblock \doi{10.1109/VR.2019.8797740}.
\newblock ISSN: 2642-5254
\bibAnnoteFile{otte_keyboards_2019}

\bibitem[{Poeschl et~al.(2013)Poeschl, Wall, and
  Doering}]{poeschl_SpatialSound_2013}
Poeschl, S., Wall, K., and Doering, N. (2013).
\newblock Integration of spatial sound in immersive virtual environments an
  experimental study on effects of spatial sound on presence.
\newblock In \emph{2013 {IEEE} {Virtual} {Reality} ({VR})}. 129--130.
\newblock \doi{10.1109/VR.2013.6549396}.
\newblock ISSN: 2375-5334
\bibAnnoteFile{poeschl_SpatialSound_2013}

\bibitem[{Raghuram et~al.(2019)Raghuram, Hill, Gibbs, and
  Maruping}]{raghuram_virtualWorkBridging_2019}
Raghuram, S., Hill, N.~S., Gibbs, J.~L., and Maruping, L.~M. (2019).
\newblock Virtual {Work}: {Bridging} {Research} {Clusters}.
\newblock \emph{Academy of Management Annals} 13, 308--341.
\newblock \doi{10.5465/annals.2017.0020}.
\newblock Publisher: Academy of Management
\bibAnnoteFile{raghuram_virtualWorkBridging_2019}

\bibitem[{Rogers et~al.(2018)Rogers, Masoodian, and
  Apperley}]{rogers2018.virtualCocktailPartyWordclouds}
Rogers, B., Masoodian, M., and Apperley, M. (2018).
\newblock A virtual cocktail party: supporting informal social interactions in
  a virtual conference.
\newblock In \emph{Proceedings of the 2018 {International} {Conference} on
  {Advanced} {Visual} {Interfaces}} (New York, NY, USA: Association for
  Computing Machinery), {AVI} '18, 1--3.
\newblock \doi{10.1145/3206505.3206569}
\bibAnnoteFile{rogers2018.virtualCocktailPartyWordclouds}

\bibitem[{Roth et~al.(2019)Roth, Bente, Kullmann, Mal, Purps, Vogeley
  et~al.}]{roth_SocialAugmentations_2019}
Roth, D., Bente, G., Kullmann, P., Mal, D., Purps, C.~F., Vogeley, K., et~al.
  (2019).
\newblock Technologies for {Social} {Augmentations} in {User}-{Embodied}
  {Virtual} {Reality}.
\newblock In \emph{25th {ACM} {Symposium} on {Virtual} {Reality} {Software} and
  {Technology}} (New York, NY, USA: Association for Computing Machinery),
  {VRST} '19, 1--12.
\newblock \doi{10.1145/3359996.3364269}
\bibAnnoteFile{roth_SocialAugmentations_2019}

\bibitem[{Roth et~al.(2018)Roth, Klelnbeck, Feigl, Mutschler, and
  Latoschik}]{roth_beyondReplication_2018}
Roth, D., Klelnbeck, C., Feigl, T., Mutschler, C., and Latoschik, M.~E. (2018).
\newblock Beyond {Replication}: {Augmenting} {Social} {Behaviors} in
  {Multi}-{User} {Virtual} {Realities}.
\newblock In \emph{2018 {IEEE} {Conference} on {Virtual} {Reality} and {3D}
  {User} {Interfaces} ({VR})}. 215--222.
\newblock \doi{10.1109/VR.2018.8447550}
\bibAnnoteFile{roth_beyondReplication_2018}

\bibitem[{Rzeszewski and Evans(2020)}]{rzeszewski2020.VRChatQuarantine}
Rzeszewski, M. and Evans, L. (2020).
\newblock Virtual place during quarantine – a curious case of {VRChat}.
\newblock \emph{Rozwój Regionalny i Polityka Regionalna} ,
  57--75\doi{10.14746/rrpr.2020.51.06}
\bibAnnoteFile{rzeszewski2020.VRChatQuarantine}

\bibitem[{Saatçi et~al.(2020)Saatçi, Akyüz, Rintel, and
  Klokmose}]{saatci_reconfiguring_2020}
Saatçi, B., Akyüz, K., Rintel, S., and Klokmose, C.~N. (2020).
\newblock ({Re}){Configuring} {Hybrid} {Meetings}: {Moving} from
  {User}-{Centered} {Design} to {Meeting}-{Centered} {Design}.
\newblock \emph{Computer Supported Cooperative Work (CSCW)} 29, 769--794.
\newblock \doi{10.1007/s10606-020-09385-x}
\bibAnnoteFile{saatci_reconfiguring_2020}

\bibitem[{Samrose et~al.(2021)Samrose, McDuff, Sim, Suh, Rowan, Hernandez
  et~al.}]{Samrose2021.MeetingCoach}
Samrose, S., McDuff, D., Sim, R., Suh, J., Rowan, K., Hernandez, J., et~al.
  (2021).
\newblock Meetingcoach: An intelligent dashboard for supporting effective \&
  inclusive meetings.
\newblock In \emph{Proceedings of the 2021 CHI Conference on Human Factors in
  Computing Systems} (New York, NY, USA: Association for Computing Machinery),
  CHI '21.
\newblock \doi{10.1145/3411764.3445615}
\bibAnnoteFile{Samrose2021.MeetingCoach}

\bibitem[{Samrose et~al.(2018)Samrose, Zhao, White, Li, Nova, Lu
  et~al.}]{Samrose2018.CollaborationCoach}
Samrose, S., Zhao, R., White, J., Li, V., Nova, L., Lu, Y., et~al. (2018).
\newblock Coco: Collaboration coach for understanding team dynamics during
  video conferencing.
\newblock \emph{Proc. ACM Interact. Mob. Wearable Ubiquitous Technol.} 1.
\newblock \doi{10.1145/3161186}
\bibAnnoteFile{Samrose2018.CollaborationCoach}

\bibitem[{Sarkar et~al.(2021)Sarkar, Rintel, Borowiec, Bergmann, Gillett, Bragg
  et~al.}]{sarkar2021.VideoMeetingChats}
Sarkar, A., Rintel, S., Borowiec, D., Bergmann, R., Gillett, S., Bragg, D.,
  et~al. (2021).
\newblock The promise and peril of parallel chat in video meetings for work.
\newblock In \emph{Extended {Abstracts} of the 2021 {CHI} {Conference} on
  {Human} {Factors} in {Computing} {Systems}} (New York, NY, USA: Association
  for Computing Machinery), {CHI} {EA} '21, 1--8.
\newblock \doi{10.1145/3411763.3451793}.
\newblock 00003
\bibAnnoteFile{sarkar2021.VideoMeetingChats}

\bibitem[{Scavarelli et~al.(2021)Scavarelli, Arya, and
  Teather}]{scavarelli2021.SocialLearningSpaces}
Scavarelli, A., Arya, A., and Teather, R.~J. (2021).
\newblock Virtual reality and augmented reality in social learning spaces: a
  literature review.
\newblock \emph{Virtual Reality} 25, 257--277.
\newblock \doi{10.1007/s10055-020-00444-8}
\bibAnnoteFile{scavarelli2021.SocialLearningSpaces}

\bibitem[{Schwartz et~al.(2020)Schwartz, Wei, Wang, Lombardi, Simon, Saragih
  et~al.}]{schwartz_eyes_2020}
Schwartz, G., Wei, S.-E., Wang, T.-L., Lombardi, S., Simon, T., Saragih, J.,
  et~al. (2020).
\newblock The eyes have it: an integrated eye and face model for photorealistic
  facial animation.
\newblock \emph{ACM Transactions on Graphics} 39, 91:91:1--91:91:15.
\newblock \doi{10.1145/3386569.3392493}
\bibAnnoteFile{schwartz_eyes_2020}

\bibitem[{Shockley et~al.(2021)Shockley, Gabriel, Robertson, Rosen, Chawla,
  Ganster et~al.}]{shockley2021.ZoomFatigue}
Shockley, K.~M., Gabriel, A.~S., Robertson, D., Rosen, C.~C., Chawla, N.,
  Ganster, M.~L., et~al. (2021).
\newblock The fatiguing effects of camera use in virtual meetings: {A}
  within-person field experiment.
\newblock \emph{Journal of Applied Psychology} 106, 1137--1155.
\newblock \doi{10.1037/apl0000948}.
\newblock 00000
\bibAnnoteFile{shockley2021.ZoomFatigue}

\bibitem[{Short et~al.(1976)Short, Williams, and
  Christie}]{short1976SocialPresenceOrg}
Short, J., Williams, E., and Christie, B. (1976).
\newblock \emph{The social psychology of telecommunications} (Toronto; London;
  New York: Wiley)
\bibAnnoteFile{short1976SocialPresenceOrg}

\bibitem[{Slater and Sanchez-Vives(2016)}]{slater_enhancing_2016}
Slater, M. and Sanchez-Vives, M.~V. (2016).
\newblock Enhancing our lives with immersive virtual reality 3
\bibAnnoteFile{slater_enhancing_2016}

\bibitem[{St\r{a}hl(1999)}]{Stahl1999.ancientVRMeetings}
St\r{a}hl, O. (1999).
\newblock Meetings for real—experiences from a series of vr-based project
  meetings.
\newblock In \emph{Proceedings of the ACM Symposium on Virtual Reality Software
  and Technology} (New York, NY, USA: Association for Computing Machinery),
  VRST '99, 164–165.
\newblock \doi{10.1145/323663.323691}
\bibAnnoteFile{Stahl1999.ancientVRMeetings}

\bibitem[{Suh et~al.(2016)Suh, Shahriaree, Hekler, and
  Kientz}]{suh2016.UserBurdenScale}
Suh, H., Shahriaree, N., Hekler, E.~B., and Kientz, J.~A. (2016).
\newblock \emph{Developing and Validating the User Burden Scale: A Tool for
  Assessing User Burden in Computing Systems} (New York, NY, USA: Association
  for Computing Machinery).
\newblock 3988–3999
\bibAnnoteFile{suh2016.UserBurdenScale}

\bibitem[{Sykownik et~al.(2021)Sykownik, Graf, Zils, and
  Masuch}]{sykownik2021.ActivitiesMotivesOfUsers}
Sykownik, P., Graf, L., Zils, C., and Masuch, M. (2021).
\newblock The {Most} {Social} {Platform} {Ever}? {A} {Survey} about
  {Activities} amp; {Motives} of {Social} {VR} {Users}.
\newblock In \emph{2021 {IEEE} {Virtual} {Reality} and {3D} {User} {Interfaces}
  ({VR})}. 546--554.
\newblock \doi{10.1109/VR50410.2021.00079}.
\newblock ISSN: 2642-5254
\bibAnnoteFile{sykownik2021.ActivitiesMotivesOfUsers}

\bibitem[{Tanenbaum et~al.(2020)Tanenbaum, Hartoonian, and
  Bryan}]{tanenbaum2020.ExpressiveNonverbalCommunicationinSVR}
Tanenbaum, T.~J., Hartoonian, N., and Bryan, J. (2020).
\newblock "{How} do {I} make this thing smile?": {An} {Inventory} of
  {Expressive} {Nonverbal} {Communication} in {Commercial} {Social} {Virtual}
  {Reality} {Platforms}.
\newblock In \emph{Proceedings of the 2020 {CHI} {Conference} on {Human}
  {Factors} in {Computing} {Systems}} (Honolulu HI USA: ACM), 1--13.
\newblock \doi{10.1145/3313831.3376606}
\bibAnnoteFile{tanenbaum2020.ExpressiveNonverbalCommunicationinSVR}

\bibitem[{Torro et~al.(2021)Torro, Jalo, and Pirkkalainen}]{torro_six_2021}
Torro, O., Jalo, H., and Pirkkalainen, H. (2021).
\newblock Six reasons why virtual reality is a game-changing computing and
  communication platform for organizations 64, 48--55.
\newblock \doi{10.1145/3440868}
\bibAnnoteFile{torro_six_2021}

\bibitem[{Wigham and Chanier(2013)}]{wigham_non-Verbal-Second-Life_2013}
Wigham, C.~R. and Chanier, T. (2013).
\newblock A study of verbal and nonverbal communication in {Second} {Life} –
  the {ARCHI21} experience.
\newblock \emph{ReCALL} 25, 63--84.
\newblock \doi{10.1017/S0958344012000250}.
\newblock Publisher: Cambridge University Press
\bibAnnoteFile{wigham_non-Verbal-Second-Life_2013}

\bibitem[{Wilcox et~al.(2006)Wilcox, Allison, Elfassy, and
  Grelik}]{wilcox_personalSpaceEmotions_2006}
Wilcox, L.~M., Allison, R.~S., Elfassy, S., and Grelik, C. (2006).
\newblock Personal space in virtual reality.
\newblock \emph{ACM Transactions on Applied Perception} 3, 412--428.
\newblock \doi{10.1145/1190036.1190041}
\bibAnnoteFile{wilcox_personalSpaceEmotions_2006}

\bibitem[{Williamson et~al.(2021)Williamson, Li, Vinayagamoorthy, Shamma, and
  Cesar}]{Williamson2021.ProxemicsSocialVRWorkshop}
Williamson, J., Li, J., Vinayagamoorthy, V., Shamma, D.~A., and Cesar, P.
  (2021).
\newblock Proxemics and social interactions in an instrumented virtual reality
  workshop.
\newblock In \emph{Proceedings of the 2021 CHI Conference on Human Factors in
  Computing Systems} (New York, NY, USA: Association for Computing Machinery),
  CHI '21.
\newblock \doi{10.1145/3411764.3445729}
\bibAnnoteFile{Williamson2021.ProxemicsSocialVRWorkshop}

\bibitem[{Williamson et~al.(2022)Williamson, O'Hagan, Guerra-Gomez, Williamson,
  Cesar, and Shamma}]{williamson_ProxemicsAudio_2022}
Williamson, J.~R., O'Hagan, J., Guerra-Gomez, J.~A., Williamson, J.~H., Cesar,
  P., and Shamma, D.~A. (2022).
\newblock Digital {Proxemics}: {Designing} {Social} and {Collaborative}
  {Interaction} in {Virtual} {Environments}.
\newblock In \emph{{CHI} {Conference} on {Human} {Factors} in {Computing}
  {Systems}} (New York, NY, USA: Association for Computing Machinery), {CHI}
  '22, 1--12.
\newblock \doi{10.1145/3491102.3517594}
\bibAnnoteFile{williamson_ProxemicsAudio_2022}

\bibitem[{Yee et~al.(2007)Yee, Bailenson, Urbanek, Chang, and
  Merget}]{yee_Nonverbal-Norms_2007}
Yee, N., Bailenson, J.~N., Urbanek, M., Chang, F., and Merget, D. (2007).
\newblock The {Unbearable} {Likeness} of {Being} {Digital}: {The} {Persistence}
  of {Nonverbal} {Social} {Norms} in {Online} {Virtual} {Environments}.
\newblock \emph{CyberPsychology \& Behavior} 10, 115--121.
\newblock \doi{10.1089/cpb.2006.9984}.
\newblock Publisher: Mary Ann Liebert, Inc., publishers
\bibAnnoteFile{yee_Nonverbal-Norms_2007}

\bibitem[{Yoshimura and Borst(2021)}]{yoshimura_study_2021}
Yoshimura, A. and Borst, C.~W. (2021).
\newblock A study of class meetings in {VR}: Student experiences of attending
  lectures and of giving a project presentation 2.
\newblock \doi{10.3389/frvir.2021.648619}
\bibAnnoteFile{yoshimura_study_2021}

\bibitem[{Zibrek and McDonnell(2019)}]{zibrek2019.PhotorealismSocialPresence}
Zibrek, K. and McDonnell, R. (2019).
\newblock Social presence and place illusion are affected by photorealism in
  embodied {VR}.
\newblock In \emph{Motion, {Interaction} and {Games}} (New York, NY, USA:
  Association for Computing Machinery), {MIG} '19, 1--7.
\newblock \doi{10.1145/3359566.3360064}
\bibAnnoteFile{zibrek2019.PhotorealismSocialPresence}

\end{thebibliography}

%%% Make sure to upload the bib file along with the tex file and PDF
%%% Please see the test.bib file for some examples of references

\end{document}

\onecolumn
\firstpage{1}

\title[Supplementary Material]{{\helveticaitalic{Supplementary Material}}}

\maketitle

\section{Supplementary Data}

Supplementary Material should be uploaded separately on submission. Please include any supplementary data, figures and/or tables. All supplementary files are deposited to FigShare for permanent storage and receive a DOI.

Supplementary material is not typeset so please ensure that all information is clearly presented, the appropriate caption is included in the file and not in the manuscript, and that the style conforms to the rest of the article. To avoid discrepancies between the published article and the supplementary material, please do not add the title, author list, affiliations or correspondence in the supplementary files.

\section{Supplementary Tables and Figures}

For more information on Supplementary Material and for details on the different file types accepted, please see \href{http://home.frontiersin.org/about/author-guidelines#SupplementaryMaterial}{the Supplementary Material section} of the Author Guidelines.

Figures, tables, and images will be published under a Creative Commons CC-BY licence and permission must be obtained for use of copyrighted material from other sources (including re-published/adapted/modified/partial figures and images from the internet). It is the responsibility of the authors to acquire the licenses, to follow any citation instructions requested by third-party rights holders, and cover any supplementary charges.

%% Figures, tables, and images will be published under a Creative Commons CC-BY licence and permission must be obtained for use of copyrighted material from other sources (including re-published/adapted/modified/partial figures and images from the internet). It is the responsibility of the authors to acquire the licenses, to follow any citation instructions requested by third-party rights holders, and cover any supplementary charges.

\subsection{Figures}

%%% There is no need for adding the file termination, as long as you indicate where the file is saved. In the examples below the files (logo1.eps and logos.eps) are in the Frontiers LaTeX folder
%%% If using *.tif files convert them to .jpg or .png
%%%  NB logo1.eps is required in the path in order to correctly compile front page header %%%

% \begin{figure}[htbp]
% \begin{center}
% \includegraphics[width=9cm]{logo1}% This is a *.eps file
% \end{center}
% \caption{ Enter the caption for your figure here.  Repeat as  necessary for each of your figures}\label{fig:1}
% \end{figure}

% \begin{figure}[htbp]
% \begin{center}
% \includegraphics[width=10cm]{logos}
% \end{center}
% \caption{This is a figure with sub figures, \textbf{(A)} is one logo, \textbf{(B)} is a different logo.}\label{fig:2}
% \end{figure}

%%% If you are submitting a figure with subfigures please combine these into one image file with part labels integrated.
%%% If you don't add the figures in the LaTeX files, please upload them when submitting the article.
%%% Frontiers will add the figures at the end of the provisional pdf automatically
%%% The use of LaTeX coding to draw Diagrams/Figures/Structures should be avoided. They should be external callouts including graphics.

%\bibliographystyle{frontiersinSCNS_ENG_HUMS} %  for Science, Engineering and Humanities and Social Sciences articles, for Humanities and Social Sciences articles please include page numbers in the in-text citations
%\bibliographystyle{frontiersinHLTH&FPHY} % for Health and Physics articles
%\bibliography{test}